\documentclass{aa}  
\usepackage{graphicx}
\usepackage{subcaption}
\usepackage{caption}
\usepackage{changepage}
\usepackage[usenames,dvipsnames]{color}
\usepackage{xcolor}
\usepackage{booktabs}
\usepackage{txfonts}
\usepackage[bookmarks, colorlinks, breaklinks, ]{hyperref} 
\usepackage{natbib}
\hypersetup{linkcolor=MidnightBlue,citecolor=MidnightBlue,filecolor=MidnightBlue,urlcolor=MidnightBlue} 
\usepackage{hyperref}
\usepackage{textcomp}
\usepackage{soul}

\usepackage{titlesec}
\usepackage{gensymb}
\usepackage{subcaption}
\usepackage{microtype}

\begin{document}

   \title{Multi-scale weak lensing detection of galaxy clusters with source redshift tomography}

   \author{L. Chappuis
          \inst{\ref{cea}} \and
          S. Pires
          \inst{\ref{cea}} \and
          G. W. Pratt
          \inst{\ref{cea}} \and
          G. Leroy 
          \inst{\ref{durham}} \and
          A. Daurelle 
          \inst{\ref{cea}} \and
          C. Giocoli
          \inst{\ref{inafoas},\ref{infnbo}} \and
          C. Carbone
          \inst{{\ref{IASFMI}}}
          }

   \institute{
    Université Paris-Saclay, Université Paris Cité, CEA, CNRS, AIM, 91190 Gif-sur-Yvette, France \label{cea}
    \email{loris.chappuis@cea.fr} 
    \and
     Institute for Computational Cosmology, Department of Physics, Durham University, South Road, Durham DH1 3LE, United Kingdom \label{durham}
    \and
     INAF -- Osservatorio di Astrofisica e Scienza dello Spazio di Bologna, Via Piero Gobetti 93/3, I-40129 Bologna, Italy \label{inafoas} \and
     INFN -- Sezione di Bologna, viale Berti Pichat 6/2, I-40127 Bologna, Italy \label{infnbo}
    \and
     INAF -- Istituto di Astrofisica Spaziale e Fisica cosmica di Milano, Via Alfonso Corti 12, I-20133 Milano, Italy \label{IASFMI}
    }
 
  \abstract
    {
    Recently, a number of methods have emerged to detect galaxy clusters solely through their weak lensing signal. Using the recently-introduced wavelet multi-scale detection method, we focus here on the potential for the use of tomographic information of the source galaxies to increase the number of weak lensing detections. We apply the $z_{s,\mathrm{min}}$-cut technique, consisting of the combination of weak lensing peak detections emerging from lensing maps obtained using different source redshift bins, to mock data sets of progressively increasing sophistication. The source redshift distribution is chosen to be $Euclid$-like, with a maximum depth of $z_{s,\mathrm{max}}=3$, and overlapping tomographic redshift bins are constructed by progressively increasing the minimum source redshift $z_{s,\mathrm{min}}$. Considering all possible detection combinations from one to four tomographic bins, we find that a single source redshift bin, with $z_{s,\mathrm{min}}=0.4$, performs as well as the combination of multiple redshift bins. By running detections on synthetic clusters of varying complexity -- from isolated Navarro Frenk White haloes to haloes embedded in and formed within N-body cosmological simulations, and considering both true and photometric source redshifts -- we show that while large-scale structure contamination and photometric redshift errors reduce the potential gains of the tomographic approach, the dominant limitation is the accumulation of spurious detections across redshift bins, leading to decreased purity at a fixed detection threshold.
    }


\maketitle

\section{Introduction}

Tracing the most prominent peaks of the matter distribution in the late Universe \citep{bardeen86}, galaxy clusters stand out as a robust cosmological probe through their mass-redshift distribution: the Halo Mass Function (HMF) \citep{Allen_2011,tinker08,despali16}. Galaxy clusters are multi-component objects that can be detected through various observational techniques. Indeed, their deep gravitational potential confines and heats primordial gas to high temperatures ($10^7$--$10^8$K). This hot ionized plasma -- the intra-cluster medium (ICM) -- accounts for most of the baryonic content in galaxy clusters (90$\%$) 
and is observable both through its X-ray emission (via thermal bremsstrahlung and line emission) and through the spectral shift it imprints on cosmic microwave background (CMB) photons via inverse-Compton scattering, the Sunyaev-Zel’dovich (SZ) effect. The remaining baryonic mass is found in the stellar component contained in the hundreds to thousands of galaxies bound to the cluster, which can be observed primarily in the optical and near-infrared wavelengths. 

These two main baryonic components -- the ICM and the member galaxies -- and their associated emission mechanisms (X-ray and tSZ for the ICM, optical for the galaxies) have been extensively exploited for cluster detection, leading to large catalogues with the statistical power required for cosmological studies \citep{bohringer_reflex_2004,  rykoff_redmapper_2014,  planckxx, bleem_spt_2015, planckxxiv,  oguri_camira_2018, hilton_act_2021, bulbul_erosita_2024}. However, baryons represent only 15$\%$ of the total cluster mass, and dark matter (DM), represents the remaining 85$\%$ \citep{Gonzalez_2013, Chiu_2018}. Detection methods relying on baryonic tracers are inherently sensitive to assumptions about baryonic physics, and robust cosmological inference from a given cluster sample therefore requires an accurate understanding of the biases introduced by the detection method \citep{Eckert_2011, Kravtsov_Borgani_2012, Pratt_2019, Wu_2022, Zhou_2024}.

In contrast, weak gravitational lensing (hereafter WL) probes the total surface mass density $\Sigma$, agnostic to the underlying composition. The bending of light-rays by local gravitational fields induces distortion of the background galaxy images (the sources) in the presence of intervening mass inhomogeneities -- the so-called gravitational lensing effect \citep{Bartelmann&Schneider_2001}. Owing to their large masses, this phenomenon can be observed around individual clusters, which act as powerful gravitational lenses \citep{Kneib_Natarajan_2011}. In the WL regime, the signal can be treated linearly and results in a small but coherent modification of the shapes (shear, $\gamma$) and sizes (convergence, $\kappa$) of the background galaxies. This subtle signal must be measured statistically on a great number of sources for each cluster \citep{Kaiser_Squires_1993}. 

A variety of approaches have been proposed to detect galaxy clusters using the WL shear or convergence. Gaussian filtering of the convergence maps allowed the recovery of several galaxy clusters with Suprime-Cam \citep{Wittman_2001, Miyazaki_2002} while the aperture-mass (AM) method \citep{Schneider_1996} opened the way to most modern techniques for WL galaxy cluster detections and was successfully applied to The Garching-Bonn Deep Survey \citep{Schirmer_2004}. This method relies on the convolution of the lensing maps with a compensated filter of a given shape to enhance contributions from matter overdensities. Various filters have emerged in the literature to optimize the expected WL signal of galaxy clusters -- by assuming, for instance, a Navarro-Frenk-White (NFW) profile -- while mitigating small and large scale noise \citep{hennawi&spergel, Maturi_2005, Hamana_2012}. 

Historically, the main limitations to increasing WL detection numbers were the source density and survey area. The arrival of the Subaru Hyper Suprime-Cam (HSC) led for the first time to catalogues of the order of a hundred WL-detected clusters. \citet{Miyazaki_2018} found 63 confirmed clusters (out of 65) over a field of 160 deg², while \citet{Hamana_2020} detected 107 confirmed clusters (out of 124) on 120 deg², and \citet{Oguri_2021} reported more than 300 clusters for 510 deg². Finally, \citet{Chiu_2024} showed cosmological constraints could be inferred through cluster abundance measurements using 129 WL-detected clusters over a 500 deg² area. With the advent of stage IV surveys such as {\it Euclid}, LSST, and Roman \citep{EuclidSkyOverview, LSST_2019, Roman_2024}, extrapolating the same techniques to their larger survey areas and higher source densities could lead to the detection of thousands of galaxy clusters. 

The present work builds upon the multi-scale WL detection algorithm for galaxy clusters presented in \citet{Leroy_2023}. Using a wavelet transform, this method decomposes the lensing convergence signal into a set of filtered maps at different spatial scales, enabling the detection of objects of different apparent sizes. We present an attempt at improving the method using the source redshift tomography. The lensing effect results in an integrated signal, where all the matter between the observer and the source contributes to the measured signal, while conversely, foreground sources (i.e. sources between the observer and a potential lens) will not be lensed. The inclusion of such foreground galaxies leads to the so-called dilution effect, which lowers the WL signal-to-noise ratio (S/N). This is a well-known problem for WL mass measurements, where it has been addressed either by treating dilution as a systematic effect \citep{Shin_2025}, or by carefully removing foreground sources from the source sample \citep{Euclid_Lesci_2024}. These strategies cannot be applied to WL cluster detection, where the lens redshift is not known a priori. In this context, dilution poses an even greater challenge: it reduces the WL S/N of clusters -- particularly at higher redshift -- thus lowering the number of detectable systems. 

Few attempts have been made in the literature to mitigate this issue. \citet{hennawi&spergel} developed a tomographic matched-filtering technique in which candidate clusters were searched for in a likelihood map parameterized by the lens redshift $z_l$; the redshift maximizing the likelihood then served as an estimate of the lens redshift. This approach yielded a 76\% increase in detections compared to a non-tomographic approach. Another strategy consists of excluding sources below a given redshift threshold $z_{s,\mathrm{min}}$ prior to constructing the WL mass map, thereby eliminating dilution for clusters with $z_l < z_{s,\mathrm{min}}$ at the cost of some information loss \citep{Chiu_2024, Trobbiani_2025}. Different values of $z_{s,\mathrm{min}}$ can be used to produce multiple WL mass maps. Peaks can then be independently detected in each map before being recombined to recover a unified set of detections. Dilution is alleviated by this progressive and discrete foreground removal, while keeping a maximum of background galaxies for low redshift lenses. This method was introduced by \citet{Hamana_2020} with the following redshift cut values $z_{s,\mathrm{min}}=[0, 0.2, 0.3, 0.4, 0.5, 0.6]$. While the individual maps yielded between 68 and 75 detections each, their combination produced 124 unique clusters -- an improvement comparable to the 76\% gain reported by \citet{hennawi&spergel}. 

This approach, combining different $z_{s, \mathrm{min}}$ to create a series of tomographic mass maps, seems particularly relevant in the case of multi-scale detection, since we expect the apparent size of clusters to vary with their redshift. Our work focuses on the potential improvements the $z_{s,\mathrm{min}}$-cut tomographic technique can bring to the multi-scale detection algorithm presented in \citet{Leroy_2023}. For comparison, we also apply the $z_{s,\mathrm{min}}$-cut to single-scale detection methods. To evaluate the performance of our method, we used a set of N-body simulations as well as a set of mock convergence fields under the spherical NFW assumption.

This paper is organised as follows. In Sect.~\ref{sec:lensing} we summarize the theory of weak lensing relevant to our work. Section~\ref{sec:data_set} describes the synthetic data set we will use throughout the paper. The detection algorithm (filters, matching method, tomographic binning schemes) is detailed in Sect.~\ref{sec:methods}. The results and associated discussion are respectively given in Sect.~\ref{sec:results} and Sect.~\ref{sec:discussion}. Finally, we give our conclusions in Sect.~\ref{sec:conclusion}. The cosmology used throughout this work comes from the \citet{Planck_2016} results. We use the common definitions of overdensity masses and radii, $M_{\Delta\mathrm{,c}}$ and $R_{\Delta\mathrm{,c}}$, as the quantities within a spherical overdensity enclosing $\Delta$ times the critical density of the Universe at the halo redshift, $\rho_{\mathrm{c}}(z) = 3 H^2(z)/(8 \pi G)$. Unless stated otherwise, we will use $\Delta=200$.

\section{Weak lensing formalism}
\label{sec:lensing}
Local inhomogeneities in the matter distribution generate gravitational fields that bend the trajectories of light rays, producing an apparent deflection angle $\alpha$. In the context of galaxy cluster WL, it is usually assumed that this deflection arises solely from the cluster under consideration. Under this assumption, the thin-lens approximation can be applied, since the physical extent of a galaxy cluster is negligible compared to the total distance travelled by light between the background source and the observer. For a more complete review, we refer the reader to \citet{Bartelmann&Schneider_2001, Umetsu_review_2020}.

\subsection{General concepts}
Geometrically, by defining the optical axis as the line between the observer and the centre of the lens, this approximation leads to the lens equation:
\begin{equation}
\beta = \theta - \alpha,
\label{eq:lens_equation}
\end{equation}
where $\theta$ is the angle between the optical axis and the observed position of the source, and $\beta$ is the angle between the optical axis and the true source position.

The gravitational potential (i.e. metric perturbation) $\Psi$ generated by the cluster gives rise to the effective lensing potential $\psi$, defined as \citep{Wright&Brainerd_2000, Umetsu_review_2020}:
\begin{equation}
    \psi(\boldsymbol{\theta}) = \frac{2}{c^2} \frac{D_{ls}}{D_l D_s} \int^{\chi_l + \Delta\chi/2}_{\chi_l - \Delta\chi/2} \Psi[\chi, r(\chi_l)\boldsymbol{\theta}]a\mathrm{d}\chi,
    \label{eq:lensing_potential}
\end{equation}
where $D_s$, $D_l$, and $D_{ls}$ are the angular diameter distances from the observer to the source, the observer to the lens, and the lens to the source, respectively. Here, $c$ is the speed of light in vacuum, $\chi$ the comoving distance, $a$ the scale factor defined as $a=1/(1+z)$, and $\Psi$ depends on both the line-of-sight distance $\chi$ and the projected distance in the lens plane $r(\chi_l)\boldsymbol{\theta}$.

The apparent deflection angle is given by the gradient of the effective lensing potential:
\begin{equation}
\alpha(\boldsymbol{\theta}) = \boldsymbol{\nabla}_\theta \psi(\boldsymbol{\theta}),
\label{eq:defl_angle_potential}
\end{equation}
so that the lens equation becomes:
\begin{equation}
\boldsymbol{\beta} = \boldsymbol{\theta} - \boldsymbol{\nabla}_{\theta}\psi(\boldsymbol{\theta}).
\label{eq:lens_eq_potential}
\end{equation}

The effective lensing properties are described by the Jacobian matrix $\mathcal{A}$ of the lens mapping. Using Cartesian coordinates $\boldsymbol{\theta} = (\theta_1, \theta_2)$, the Kronecker delta $\delta_{ij}$, and the notation $\psi_{ij} = \frac{\partial^2{\psi(\boldsymbol{\theta})}}{\partial{\theta_i}\partial{\theta_j}}$, it is expressed as:
\begin{equation}
    \mathcal{A}(\boldsymbol{\theta}) = \frac{\partial{\boldsymbol{\beta}}}{\partial{\boldsymbol{\theta}}} = \delta_{ij} - \psi_{ij} =
    \begin{pmatrix}
        1 - \kappa - \gamma_1 & -\gamma_2\\
        -\gamma_2 & 1 - \kappa + \gamma_1
    \end{pmatrix}.
\label{eq:jacobian}
\end{equation}

From Eq.~\ref{eq:jacobian}, two main observable effects can be identified: convergence ($\kappa$), and shear (given by its two components, $\gamma_1$ and $\gamma_2$).
The convergence modifies the trace of the Jacobian, leading to changes in the apparent size (magnification) of the source:
\begin{equation}
\kappa = \frac{1}{2}\left( \psi_{11} + \psi_{22}\right) = \frac{1}{2}\Delta\psi,
\label{eq:kappa_nabla_potential}
\end{equation}
The complex shear $\gamma$ induces an apparent axial distortion of the source:
\begin{equation}
\begin{aligned}
    \gamma &= \gamma_1 + i \gamma_2,\\
    \gamma_1 &= \frac{1}{2}\left( \psi_{11} - \psi_{22} \right),\\
    \gamma_2 &= \psi_{12}.
\end{aligned}    
\label{eq:shear_complex}
\end{equation}

To connect the convergence to the projected matter distribution, we now introduce the surface mass density. In the thin-lens approximation, the three-dimensional mass density $\rho$ of the lens is integrated along the line of sight, yielding
\begin{equation}
\Sigma(\boldsymbol{\theta}) = \int_{0}^{\chi_s} \left( \rho(\chi, r(\chi)\boldsymbol{\theta})\right) a\, \mathrm{d}\chi,
\label{eq:sigma}
\end{equation}
where $\chi_s$ is the comoving distance to the source.

This projected mass density is directly related to the lensing potential. Indeed, Eq.~\ref{eq:kappa_nabla_potential} can be interpreted as the two-dimensional equivalent of Poisson's equation of the gravitational potential, and thus:
\begin{equation}
\Delta \psi(\boldsymbol{\theta}) = 2 \frac{D_{ls} D_l}{D_s} \frac{4 \pi G}{c^2} \Sigma(\boldsymbol{\theta}),
\label{eq:potential_sigma}
\end{equation}
leads to the familiar expression:
\begin{equation}
\kappa(\boldsymbol{\theta}) = \frac{\Sigma(\boldsymbol{\theta})}{\Sigma_{\mathrm{crit}}},
\label{eq:kappa_sig_sigcrit}
\end{equation}
where the critical surface mass density is defined as
\begin{equation}
\Sigma_{\mathrm{crit}} = \frac{c^2}{4\pi G} \frac{D_s}{D_l D_{ls}}.
\end{equation}

The strength of the lensing signal depends on the relative lens-source geometry, encoded in the distance ratio $\frac{D_{ls}}{D_s}$. It is convenient to describe this geometric dependence through a lensing efficiency -- or lensing kernel -- which weights the source contribution to the observed signal according to the redshift: $w(z_l,z_s) \equiv \Sigma_{\mathrm{crit}}^{-1}$. For a realistic source population, the effective lensing response must account for the source redshift distribution $p(z_s)$ (see Appendix \ref{sec:appendix:S/N}).

The unitless convergence field $\kappa$ therefore provides a direct link between the observed lensing signal and the projected matter distribution of the lens, normalised by the geometric factor $\Sigma_{\mathrm{crit}}$ that encodes the source–lens–observer configuration.
The averaged ellipticity of the background galaxies provides an estimator of the WL effect, the reduced shear $g$, which is linked to the shear and convergence through the following relation:
\begin{equation}
    g(\boldsymbol{\theta}) = \frac{\gamma(\boldsymbol{\theta})}{1-\kappa(\boldsymbol{\theta})}.
\label{eq:reduced_shear}
\end{equation}

\subsection{Mass inversion}
The convergence field, which is not a direct observable, can be recovered through mass inversion methods, which exploit the link between shear and convergence to recover the latter. We present here a brief overview of the widely used formalism described by \citet[][hereafter KS93]{Kaiser_Squires_1993}. Using the complex differential operators
\[
\partial \equiv \partial_1 + \mathrm{i}\partial_2, 
\qquad 
\partial^* \equiv \partial_1 - \mathrm{i}\partial_2,
\]
the convergence and shear can be expressed in terms of the lensing potential $\psi$ as
\begin{align}
\kappa(\boldsymbol{\theta}) &= \tfrac{1}{2}\,\partial^* \partial \,\psi(\boldsymbol{\theta}),
\label{eq:kappa_spin}\\
\gamma(\boldsymbol{\theta}) &= \tfrac{1}{2}\,\partial \partial \,\psi(\boldsymbol{\theta}).
\label{eq:gamma_spin}
\end{align}
These relations imply an analytic connection between $\kappa$ and $\gamma$, which becomes
particularly transparent in Fourier space. Denoting Fourier transforms with a hat,
and writing the wavevector as $\boldsymbol{k}=(k_1,k_2)$, one obtains
\begin{equation}
    \hat{\gamma}(\boldsymbol{k}) = \hat{P}(\boldsymbol{k})\, \hat{\kappa}(\boldsymbol{k}),
\end{equation}
with
\begin{equation}
\begin{aligned}
    \hat{P}(\boldsymbol{k}) &= \hat{P}_1(\boldsymbol{k}) + i\,\hat{P}_2(\boldsymbol{k}), \\
    \hat{P}_1(\boldsymbol{k}) &= \frac{k_1^2 - k_2^2}{k_1^2 + k_2^2}, \\
    \hat{P}_2(\boldsymbol{k}) &= \frac{2\,k_1 k_2}{k_1^2 + k_2^2}.
\end{aligned}
\end{equation}
Inverting gives the KS93 estimator,
\begin{equation}
    \hat{\kappa}(\boldsymbol{k}) = \hat{P}^*(\boldsymbol{k})\, \hat{\gamma}(\boldsymbol{k}),
\end{equation}
which can be separated into E- and B-modes:
\begin{equation}
\begin{aligned}
    \hat{\kappa}_E(\boldsymbol{k}) &= \hat{P}_1 \,\hat{\gamma}_1 + \hat{P}_2 \,\hat{\gamma}_2, \\
    \hat{\kappa}_B(\boldsymbol{k}) &= -\hat{P}_2 \,\hat{\gamma}_1 + \hat{P}_1 \,\hat{\gamma}_2,
\end{aligned}
\end{equation}
where $\hat{\gamma}_1$ and $\hat{\gamma}_2$ are the Fourier transforms of the shear components.

However, this inversion is undefined at $\boldsymbol{k} = 0$, corresponding to a constant offset in $\kappa$. This reflects the mass-sheet degeneracy, which arises because shear depends only on derivatives of the potential and is therefore insensitive to a uniform mass layer. 

\subsection{Aperture mass}
By selecting a filter $U(\theta)$ targeting a specific scale, the aperture mass (AM) formalism \citep{Schneider_1996, Schneider_1998} was introduced as a method to identify massive structures from shear measurements. It defines a scalar quantity, the aperture mass, as a filtered projection of the convergence field:
\begin{equation}
    M_{\mathrm{ap}}(\boldsymbol{\theta}_0) = \int \mathrm{d}^2\theta \, U(|\boldsymbol{\theta} - \boldsymbol{\theta}_0|) \ \kappa(\boldsymbol{\theta}),
    \label{eq:AM}
\end{equation}
where the compensation condition,$\int \theta U(\theta)=0$, ensures that $M_{\mathrm{ap}}$ is locally insensitive to the additive component of the mass-sheet degeneracy. 
High values of $M_{\mathrm{ap}}$ then correspond to significant mass concentrations. The AM technique thus provides a direct, scale-dependent approach to WL detection of galaxy clusters and forms the conceptual basis for subsequent refinements of the filter function -- either through model-dependent optimal filtering \citep{Maturi_2005, Trobbiani_2025} or through multi-scale analyses \citep{Leroy_2023}, as adopted in this work.

\section{Simulated dataset}
\label{sec:data_set}
\subsection{N-body simulation}
The simulations used in this work are two realisations from the DEMNUni Covariance project (DEMNUni-Cov; \citealt{parimbelli_21, baratta_22, gouyou_25, Ingoglia_etal_2025, Bel_etal_2025}), a subset of the Dark Energy and Massive Neutrino Universe (DEMNUni; \citealt{Carbone_2016, Parimbelli_2022}) simulation suite, adopting standard $\Lambda$CDM cosmological parameters from \citet{Planck_2016}. Initial conditions were generated at $z=99$ using a theoretical linear power spectrum computed with \texttt{CAMB} \citep{Lewis_2000}. Each simulation follows the gravitational evolution of $1024^3$ cold dark matter particles in a cubic volume of side length $1 \, h^{-1} \, \mathrm{Gpc}$, corresponding to a particle mass resolution of $m_{\rm p} \approx 8 \times 10^{10} \, h^{-1} \, M_\odot$. A total of 63 snapshots were produced, spanning redshifts from $z=99$ to $z=0$.

Dark matter haloes were identified in each snapshot using a Friends-of-Friends (FoF) algorithm \citep{Davis_1985} with linking length $\lambda = 0.2d$, where $d$ is the mean particle separation. Gravitationally bound structures were then extracted with the {\tt SUBFIND} algorithm \citep{Springel_2001}, providing the redshift, mass ($M_{200\mathrm{,c}}$), and radius ($R_{200\mathrm{c}}$) of each halo. 

Weak lensing past light-cones were constructed from $z=0$ to $z=4$ using the \texttt{MapSim} routine \citep{Giocoli_2014}. The Born approximation \citep{Bartelmann&Schneider_2001} was adopted, neglecting lens-lens coupling, which is known to provide an accurate description of weak cosmic shear \citep{Schafer_2012}. Galaxy catalogues were then produced from these maps, containing randomly distributed positions (RA, Dec), shear components, and redshifts. The redshift distribution of the galaxies was derived from COSMOS2015 photometric redshifts \citep{Laigle_2016}, applying a cut at $i_{\rm E} \leq 24.5$ to mimic the expected {\it Euclid} $n(z)$, as described in \citet{Euclid_Ajani_2023}. 

To emulate observational systematics, additional sources of uncertainty were included in the galaxy catalogues. Shape noise, arising from the intrinsic ellipticity distribution of galaxies and measurement errors, was modelled as Gaussian noise with zero mean and dispersion $\sigma_\epsilon = 0.26$ per shear component \citep{Leauthaud_2007, Schrabback_2018, Euclid_Martinet_2019}. Photometric redshift errors were implemented with a scatter of $\sigma_z = 0.05(1+z)$, with 10$\%$ of redshifts flagged as catastrophic outliers. 
 Each simulated field covers $10^\circ \times 10^\circ$ and contains approximately $10^7$ galaxies, corresponding to a density of $n_g \simeq 30 \, \mathrm{arcmin}^{-2}$.

In this work, we used 12 such simulated fields, corresponding to a total area of 1200 $\mathrm{deg}^2$. These fields were generated from two independent DEMNUni-Cov realisations, each providing six randomised light-cones. For each field, both a galaxy catalogue and a halo catalogue were produced. In the following, we refer to these as realisations of the simulated fields, to distinguish them from the underlying DEMNUni-Cov cosmological realisations on which they are based.

Hereafter, when we mention the so-called simulation cluster catalogue, we refer to the DM haloes identified by the FoF algorithm, for which the WL S/N is expected to be greater than two. For our $Euclid$-like source density and redshift distribution, this condition is respected for haloes above the following mass-redshift cut \citep{Andreon_2012}:
\begin{equation}
    \log_{10}\left(\frac{M_{200c}}{M_\odot}\right) > 13.3 + 1.049\,z + 0.489\,z^{2}\;.
    \label{eq:diagonal_cut}
\end{equation}  
We show the cut in the mass-redshift space and the redshift distribution of the catalogue haloes after its application in Fig.~\ref{fig:fields} (left and right panels, respectively). To assess the performance of the detection algorithm, we matched the detected peaks to the DM haloes in the catalogue. Excluding objects expected to have a low S/N from the matching catalogue reduces the likelihood of associating a spurious detection with a real halo.  

\subsection{NFW mocks}
\subsubsection{NFW mocks without LSS}
In addition to the N-body simulation presented above, we also produced simpler semi-analytical mocks under the assumption that all matter inhomogeneities in the field are produced by DM haloes with spherically symmetric Navarro–Frenk–White (NFW, \citealt{Navarro_1997}) density profiles. We populated $10^\circ \times 10^\circ$ fields with NFW haloes sharing the same mass, redshift, and RA, Dec positions, as the clusters of the N-body simulation catalogues above the mass-redshift cut presented in Eq.~\ref{eq:diagonal_cut}.

\begin{figure*}[htbp]
    \centering
    \includegraphics[width=\linewidth, trim={3cm 1.8cm 2cm 2cm}, clip]{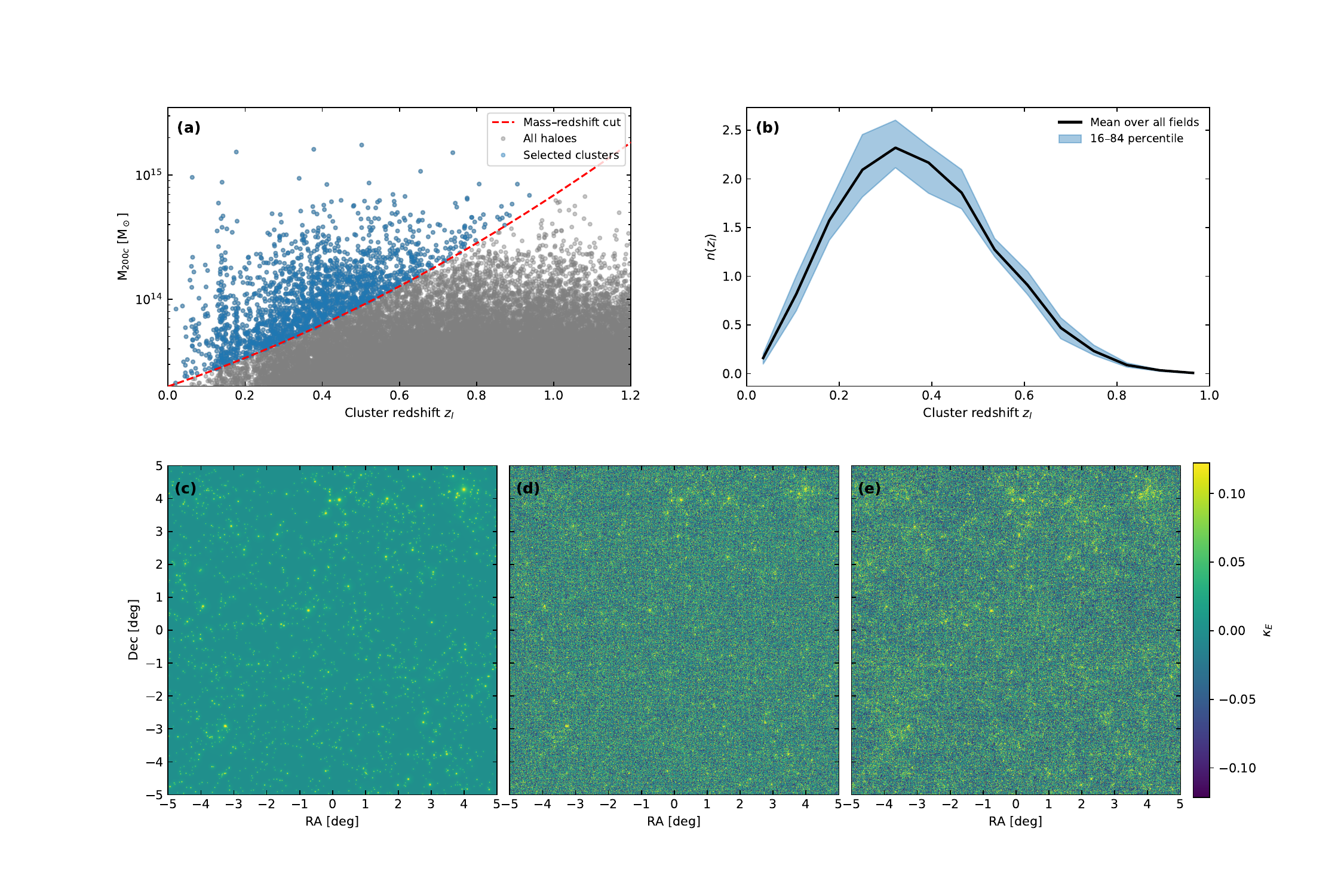}
    \caption{
    (a) Mass–redshift ($M_{200\mathrm{c}}$–$z_l$) plane for one simulated field, showing all haloes in the simulation catalogue (grey dots). Haloes selected by the mass–redshift cut defined in Eq.~\ref{eq:diagonal_cut} (red dashed line) are highlighted in blue.
    (b) Redshift distribution of the clusters selected by the mass–redshift cut, shown as the mean over all simulated fields (black line), with the 16th–84th percentile range indicated by the blue shaded region.
    (c) Mock $\kappa_E$ convergence field populated by NFW haloes with mass, redshift, and angular positions from the selected halo catalogue.
    (d) Same mock convergence field, with added shape noise.
    (e) Original N-body $\kappa_E$ from which the halo catalogue was extracted.
    }
    \label{fig:fields}
\end{figure*}

We proceeded by slicing the lens redshift $z_l$ distribution to produce projected matter maps that only account for the lenses in a given redshift slice. Similarly, we produced maps using source redshift distributions with different minimum redshift cuts $z_{s, \mathrm{min}}$. Noiseless mock convergence maps were then generated by projecting the injected NFW haloes, as presented in Fig.~\ref{fig:fields}c. The details of the methodology used to go from the 3D NFW profiles to the convergence while accounting for the halo redshift and the redshift of the source distribution are given in Appendix~\ref{sec:appendix:nfw_mocks}. Gaussian noise was injected into each field according to the shape noise and source density. The resulting noisy convergence maps were then shifted to have zero mean, as expected from the properties of real reconstructed fields. We display one of the final NFW mock convergence maps in Fig.~\ref{fig:fields}d, compared to the initial N-body convergence field from which the halo population was drawn in Fig.~\ref{fig:fields}e.

These simplified convergence fields, compared to those derived from full N-body simulations, allow us to evaluate the performance of our detection algorithm in the absence of correlated LSS noise and satellite haloes, as we model only the haloes above the mass-redshift cut. This enables us to assess whether the detected peaks correspond to genuine halo positions without signal boosting from LSS line-of-sight alignments, and without depending on the accuracy of the FoF algorithm. 

\subsubsection{NFW injection in the N-body fields}

The purpose of this injection procedure was to provide an intermediate level of realism between the idealised NFW mocks devoid of LSS signal and the fully self-consistent N-body simulation fields. Specifically, the injected fields retain the realistic uncorrelated LSS fluctuations and projection effects present in the N-body convergence maps, while allowing the properties of the injected haloes to be fully controlled.

Using the same method as for the LSS-free mocks, we injected NFW haloes in the convergence fields derived from the N-body simulation. To preserve the HMF of the N-body simulation (also used in the LSS-free mocks), we randomly drew mass and redshift from the haloes of the N-body simulation catalogue above the mass-redshift cut defined in Eq.~\ref{eq:diagonal_cut}. The angular positions (RA, Dec) of the injected haloes were randomised across the field to prevent a systematic superposition with the existing haloes. 

For each realisation, 200 such haloes were injected. This procedure was undertaken one halo at a time, to ensure that the injected signal remains a local perturbation of the convergence field and does not significantly alter its statistical properties. This way, our detection algorithm can be applied after each injection without biasing the process.

With this approach, the performance of the detection algorithm can be evaluated in three complementary regimes of increasing physical complexity: (i) isolated NFW haloes in the absence of LSS, (ii) NFW haloes injected in realistic LSS fluctuations, and (iii) haloes self-consistently formed in the N-body simulation. The injected fields will therefore allow us to isolate the impact of LSS noise on the detection efficiency, separated from the properties of the individual haloes (deviations from NFW, cluster morphology, and correlated matter) present in the N-body simulations.

\section{Methods}
\label{sec:methods}
\subsection{The multi-scale detection algorithm}

The multi-scale maps obtained by applying the wavelet transform to the WL convergence field have been shown to be formally equivalent to AM filtering at the corresponding scales \citep{Leonard_2012}, with the additional advantage of offering a set of functions localised in both real and Fourier space. This, along with the wide range of apparent cluster sizes owing to their mass and redshift distribution, motivates the use of the wavelet transform for WL cluster detection. Following \citet{Leroy_2023} -- towards which we refer the reader for a more complete description -- we used the starlet transform, an isotropic undecimated wavelet transform \citep{Starck_1998, Starck_2006}. The convergence signal can then be decomposed into multi-scale wavelet maps $W_i(i=1,...,J)$ as follows:
\begin{equation}
\kappa(\theta) = C_J(\theta) + \sum_{i=1}^{J} W_i(\theta),
\label{eq:starlet}
\end{equation}
where $C_J$ is the smoothed version of the convergence for $J$ decomposition scales.
This decomposition is strictly equivalent to an AM filtering where each wavelet scale $i$ corresponds to the following filter \citep{Leonard_2012}:
\begin{equation}
    \begin{aligned} & U_{W_i}(u)=\frac{1}{9}\left[93|u|^3-64\left(\left|\frac{1}{2}-u\right|^3+\left(\frac{1}{2}+u\right)^3\right)\right. \\ & \left.\quad+18\left(|1-u|^3+|1+u|^3\right)-\frac{1}{2}\left(|2-u|^3+|2+u|^3\right)\right].\end{aligned}
\label{eq:wvlt_filters}
\end{equation}
Here, $u=\frac{\theta}{2^i p}$, highlights the dyadic behaviour of the starlet transform, where each wavelet scale $W_i$ targets features with an  angular radius of $2^{i-1}$ times the pixel size $p$. 

In our study, the 10°$\times$10° shear catalogues were pixelated into 1024$\times$1024 pixels, resulting in a pixel size of approximately 0.59 arcmin. The convergence maps (both $E-$ and $B-$modes) were then reconstructed using the KS93 algorithm \citep{Kaiser_Squires_1993}. Finally, the starlet transform was applied with $J=7$ scales. Scales with $i\in[2,3,4]$ were used as detection channels and are referred to as W2, W3, and W4, , respectively. These scales target objects of radii $\approx$ 1.17, 2.34, and 4.69 arcmin (that is, $2\times p$, $4\times p$, $8\times p$).

\subsection{Improvements since \citet{Leroy_2023}}

\subsubsection{Multi-scale thresholding and peak detection}

Pixels in the W2, W3, and W4 maps with values higher than their eight direct neighbours were identified as peaks. A detection threshold has to then be defined, for which we used the $B-$mode convergence maps, which give us access to the noise distribution. Indeed, since the lensing signal arises from a scalar potential, it is expected to generate only $E-$modes. We defined a scale-dependent detection threshold to maintain a constant false detection rate for each scale. This was achieved by imposing a threshold that maintains the number of detected B-mode (noise) peaks unchanged.
However, due to the dyadic nature of the starlet transform, higher order wavelets (corresponding to larger spatial scales) introduce correlations between neighbouring pixels. This scale-dependent loss of resolution must be accounted for to ensure that the thresholding maintains a fixed fraction of false detections across different scales.

Let the threshold at wavelet scale $i$ be expressed as:
\begin{equation}
    \alpha_i = k_i\sigma_i
\end{equation}
where $\sigma_i$ is the noise dispersion at the scale $i$ and $k_i$ is a dimensionless coefficient. The probability $p_i$ that a noise fluctuation exceeds this threshold can then be derived from the cumulative distribution function (CDF) of the noise distribution $F_{\kappa_B}$:
\begin{equation}
    p_i = 1 - F_{\kappa_B}(k_i).
\end{equation}
To preserve a constant expected number of false detections when moving from scale $i$ to $i+1$, we therefore require that the fraction of pixels exceeding the threshold scales inversely with the number of statistically independent elements in the map.
Assuming a dyadic decomposition, the number of independent pixels decreases by a factor of four between successive wavelet scales.
Hence, we impose
\begin{equation}
    p_{i + 1} =  4 * p_i,
\end{equation}
leading to the corresponding threshold
\begin{equation}
\alpha_{i+1} = F_{\kappa_B}^{-1}\left(1 - p_{i + 1} \right) \sigma_{i+1},
\end{equation}
where $F_{\kappa_B}^{-1}$ is the percent point function (PPF); that is, the inverse CDF.

\subsubsection{Merging of the wavelet peak detections}
\label{sec:methods:merging}

The same overdensity can overcome the detection threshold in more than one scale, making it essential to recombine these multiple detections. This can be handled either during the matching with the catalogue, as in \citet{Leroy_2023}, or by merging the detected peaks across the different scales before undertaking the matching. Here we follow the second approach, as it is more appropriate for producing a unified detection catalogue that is agnostic to the choice of matching catalogue.

The merging was undertaken hierarchically, starting from the largest scale (W4). For each W4 detection, we searched around its centre for W3 peaks located within a radius corresponding to the W4 scale (eight pixels). If no W3 peak was found, we instead looked for W2 detections in the same region. If subdetections (W3 or W2) were found, we assessed whether they were spatially distinct by requiring a minimum separation corresponding to their particular  scale (four pixels for W3 and two pixels for W2). Distinct subdetections were retained individually; otherwise, the parent (W4) detection was kept. 

\begin{figure*}[htbp]

      \centering
      \includegraphics[width=1.\linewidth, trim={0cm 0cm 0cm 0cm},clip]{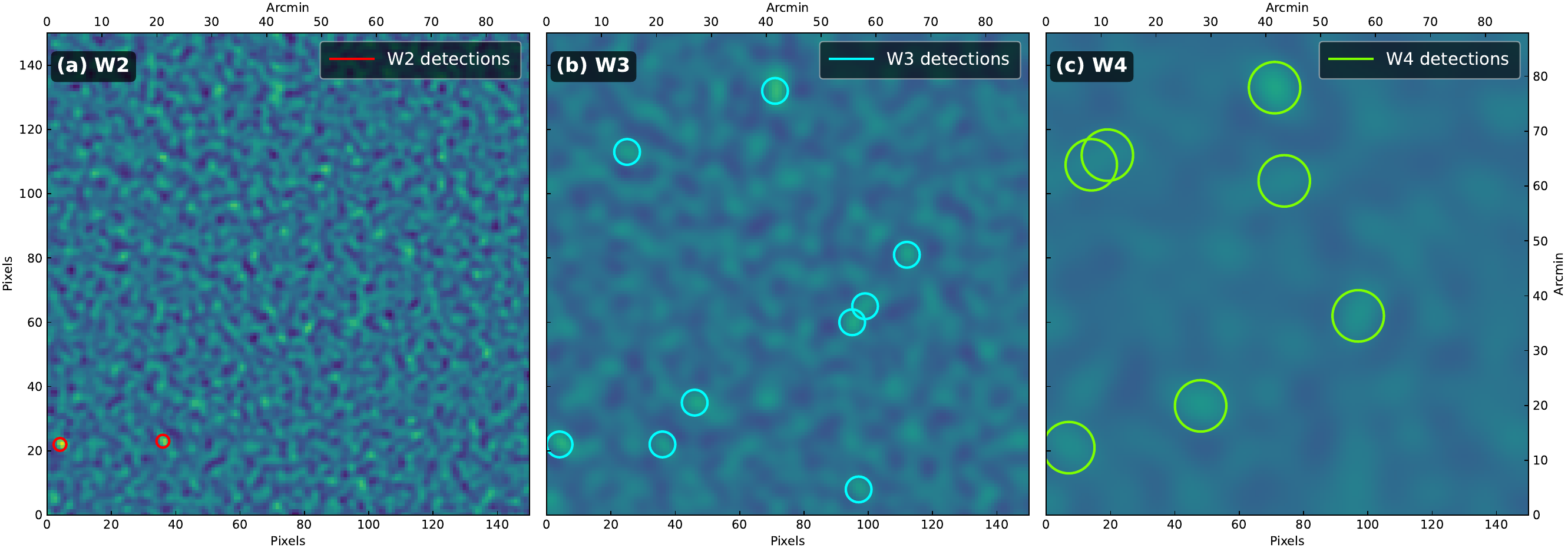}

    \caption{Multi-scale peak detections, shown on a patch of an N-body convergence field in the W2, W3, and W4 scales: respectively (a), (b), and (c)).
    Detection peaks are circled with a radius matching the typical size of the given wavelet scale (red: W2, cyan: W3, and green: W4). 
    The detection threshold was set on W2 as $4.6\sigma$, and the convergence map was generated using the full source redshift distribution (i.e. $z_{s,\mathrm{min}}=0$).}
    \label{fig:unmerged_wvlt}
\end{figure*}

\begin{figure}[htbp]

      \centering
      \includegraphics[width=1.\linewidth, trim={0cm 0.5cm 0cm 0.5cm},clip]{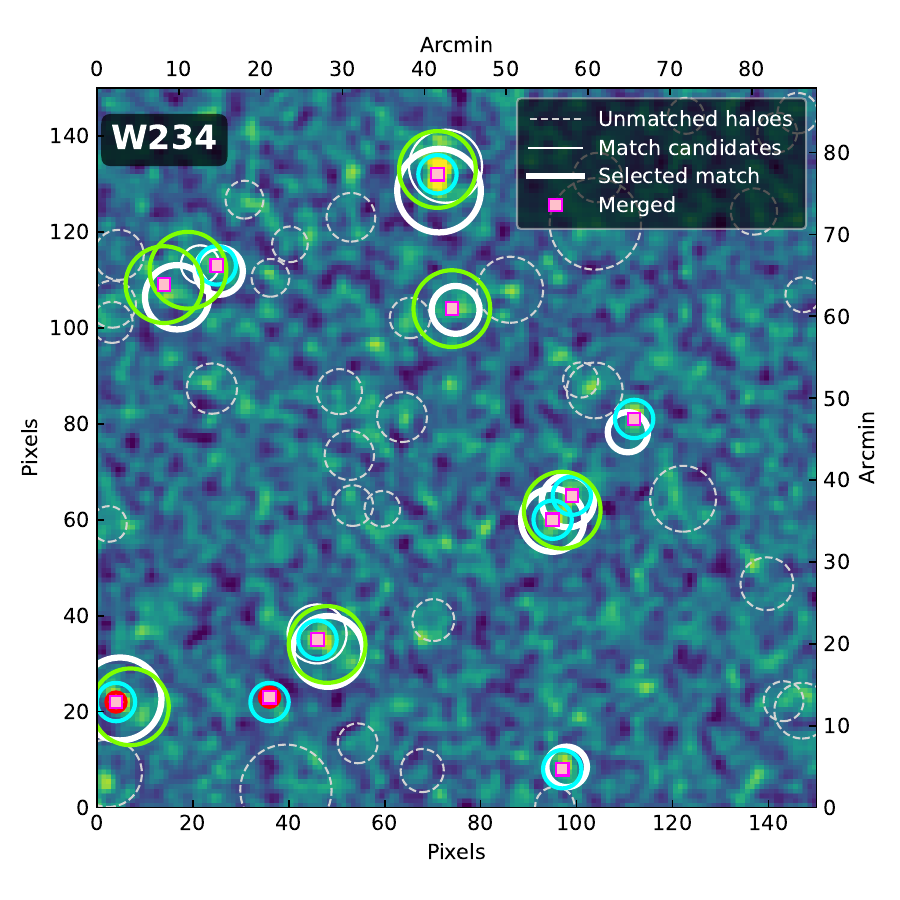}

    \caption{Merging of multi-scale detections on W234 (W2+W3+W4). Pink squares mark the final merged detection positions. Thick white circles indicate haloes matched to detections; thin white circles show haloes rejected by the S/N matching when multiple candidates fall within the matching radius. Unmatched haloes are shown as dashed grey circles. Circle sizes scale with $\theta_{200}$. The detection threshold is $4.6\sigma$ on W2, and the convergence map uses the full source redshift distribution ($z_{s,\mathrm{min}}=0$).}
    \label{fig:merged_wvlt}
\end{figure}

For each W3 detection not already associated with a W4 peak, we repeated the procedure: nearby W2 detections were grouped under the W3 detection if they were not clustered within each other. Finally, any W2 detections not already assigned to a W3 or W4 peak were added independently to the final catalogue. This procedure allows us to avoid double-counting the same overdensity detected at multiple scales, while also preserving the detection of nearby smaller-scale structures that may be blended within a larger scale overdensity. Finally, the merged detection was assigned the spatial position of its smallest detection scale. Detections on each scale are illustrated in Fig.~\ref{fig:unmerged_wvlt}, and we show the outcome of our spatial merging algorithm in Fig.~\ref{fig:merged_wvlt}.

\subsubsection{S/N-based Matching}
\label{sec:methods:matching}
As our baseline matching algorithm must set an association distance for each scale, we followed the maximum matching distance (MMD) strategy, as defined by \citet{Leroy_2023}. The MMD value is determined at each scale by examining histograms of the angular distance between each detection and its associated closest catalogue halo, $d$. More details can be found in \citet{Leroy_2023}. 

Although the MMD sets an optimal association distance for each scale, it does not directly discriminate between two potential associations that would both respect the condition $d < \mathrm{MMD}$.
To deal with this problem, \citet{Leroy_2023} undertook the matching with the simulation catalogue in two steps: 1)
    for each detection, all the haloes of the catalogue were ordered by their normalised distance to the detection. The normalised distance is defined as
    \begin{equation}
        {d_{\mathrm{norm}}} = \frac{d}{\theta_{200}},
        \label{eq:dnorm}
    \end{equation}
    where $d$ is the angular distance between the detection and the halo, and $\theta_{200}$ the apparent angular size of the halo corresponding to its physical radius $R_{200\mathrm{c}}$; 2)
    if the catalogue halo with the smallest $d_{\mathrm{norm}}$ value respects the condition $d < \mathrm{MMD}$, the halo is matched to the detection.

However, $\theta_{200}$ is not a one-to-one proxy for the lensing S/N, as it omits crucial information regarding the redshift-dependent lensing efficiency and filter shape. For this reason, we modified the association procedure to account for the theoretical S/N of the halos in the matching procedure:
    1) if a halo respects the condition $d < \mathrm{MMD}$, the halo is matched to the detection;
    2) if several haloes match this condition, the one with the highest theoretical S/N in the given wavelet mass-map is chosen for the matching.
A patch of a wavelet mass-map illustrating detections and matched haloes is illustrated in Fig.~\ref{fig:merged_wvlt}. Details of the calculation of the theoretical S/N used in the association procedure are given in Appendix \ref{sec:appendix:S/N}.

\subsection{Source redshift tomography}

\begin{figure*}[htbp]
    \centering
    \begin{subfigure}{0.49\linewidth}
      \includegraphics[width=\linewidth]{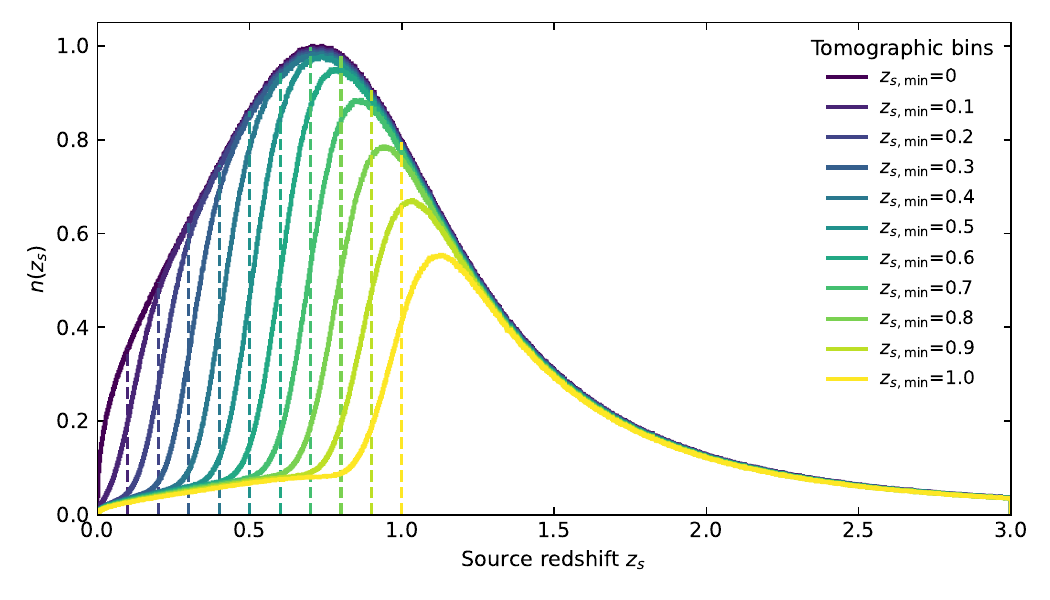}
    \end{subfigure}\hfill
    \begin{subfigure}{0.49\linewidth}
      \includegraphics[width=\linewidth]{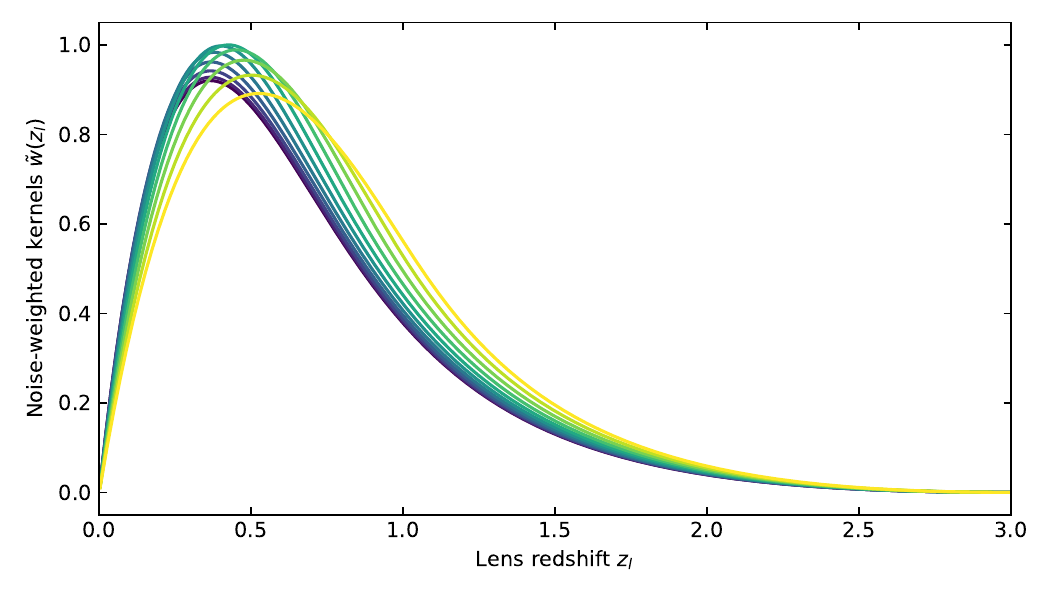}
    \end{subfigure}

    \caption{
    Left: Source redshift distributions for 
    tomographic bins defined by lower redshift cuts
    $z_s \ge z_{s,\mathrm{min}}$ (vertical dashed lines).
    Solid curves show the resulting true-redshift distributions when bins are selected using
    photometric redshifts, highlighting inter-bin leakage due to photometric redshift uncertainties. 
    Right: lensing efficiency kernels as a function of lens redshift for each source bin,
    weighted by the shape noise contribution.
    }
    \label{fig:nz_wz}
\end{figure*}

To mitigate the dilution of the WL signal by foreground sources, we adopted the $z_{s,\mathrm{min}}$-cut tomographic technique introduced by \citet{Hamana_2020}. Detections were performed independently on lensing maps constructed from source galaxy samples selected with increasing lower redshift bounds $z_{s,\mathrm{min}}$, while keeping a common upper bound $z_{s,\mathrm{max}}$, set by the depth of the survey and the reliability of photometric redshifts. We built eleven overlapping redshift bins with $z_{s, \mathrm{min}} \in [0, 0.1, 0.2, \dots, 0.9, 1]$. We show the effect of this binning on the source redshift distribution in the left hand panel of Fig.~\ref{fig:nz_wz}. The highest value of $z_{s, \mathrm{min}}$ was chosen to be $z=1$, as detection of clusters above this redshift is very unlikely due to the characteristics of the HMF and the theoretical lensing S/N of galaxy clusters  (Fig.~\ref{fig:fields}a) at these redshifts. With increasing $z_{s,\mathrm{min}}$, foreground sources are progressively removed, reducing signal dilution while shifting the peak of the lensing kernel to higher redshift, with both effects leading to higher-redshift lens detectability at the cost of a lower source density. To highlight the interplay of the lensing kernel and source density, we show the lensing kernels corresponding to each source redshift, normalized by the shape noise contribution ($\sqrt{n_g}/\sigma_\epsilon$), in the right hand panel of Fig.~\ref{fig:nz_wz}.

Here we focus on the optimisation of the redshift binning of the source galaxies in a {\it Euclid}-like survey, and on investigating potential improvements from application of this technique on cluster detection performance. For each field (either from the N-body simulations or from our semi-analytical mocks), we constructed a set of convergence maps $\kappa_i$ using source galaxy bins with $z_{s, \mathrm{min}} \in [0, 0.1, 0.2, \dots, 0.9, 1]$, resulting in eleven overlapping source samples. The wavelet transform and detection algorithm were then applied independently to each $\kappa_i$, producing a corresponding set of detection catalogues $\mathrm{det}_i$. We explored different strategies for combining the resulting tomographic information by considering configurations with 1, 2, 3, and 4 source redshift bins. In the single-bin case, detections obtained from each individual $z_{s, \mathrm{min}}$-cut (that is, each detection catalogue $\mathrm{det}_i$) were analysed independently and compared. In the multi-bin cases, we merged the different detection catalogues $\mathrm{det}_i$, generating combined catalogues from the union of detections from the selected bins where overlapping detections are merged. All possible combinations of source redshift bins were explored for a given number of bins, allowing us to empirically identify the combinations of $z_{s, \mathrm{min}}$ cuts that maximised detection performance.

Finally, we compared the results obtained for different numbers of source redshift bins in order to assess the potential gain from increasing tomographic information. As described in Sect.~\ref{sec:data_set}, we assumed a {\it Euclid}-like source redshift distribution for both the N-body simulation fields and the semi-analytic mocks, yielding $z_{s,\mathrm{max}}=3$. We additionally investigated the impact of photometric redshift uncertainties by comparing results obtained with and without photometric errors.

\subsection{Purity and completeness}

The metric used to evaluate detection performance relies on the completeness $C$ vs purity $P$ plane. These are defined as follows:
\begin{equation}
    C = \frac{n_{\mathrm{matched}}}{n_\mathrm{cat}},
    \qquad
    P = \frac{n_{\mathrm{matched}}}{n_\mathrm{det}} .
    \label{eq:completeness_purity}
\end{equation}
where $n_{\mathrm{det}}$ is the total number of detections, $n_{\mathrm{cat}}$ is the number of haloes in the catalogue used for the matching, and $n_{\mathrm{matched}}$ is the number of detections that were matched to a halo of the catalogue.
We varied the detection thresholds $k_i$ to build completeness vs. purity curves.
As we aim to compare multiple tomographic configurations, resulting in multiple completeness vs. purity curves, we will rank each method according to the area under the curve (AUC) of the completeness vs purity curves, in the purity range [0.85, 0.95].

\section{Results}
\label{sec:results}
\subsection{NFW mocks without LSS}

\subsubsection{Proof of concept with lens redshift slices}

\begin{figure*}[htbp]
    \centering
    \includegraphics[width=\linewidth, trim={0cm 0cm 0cm 0cm},clip]{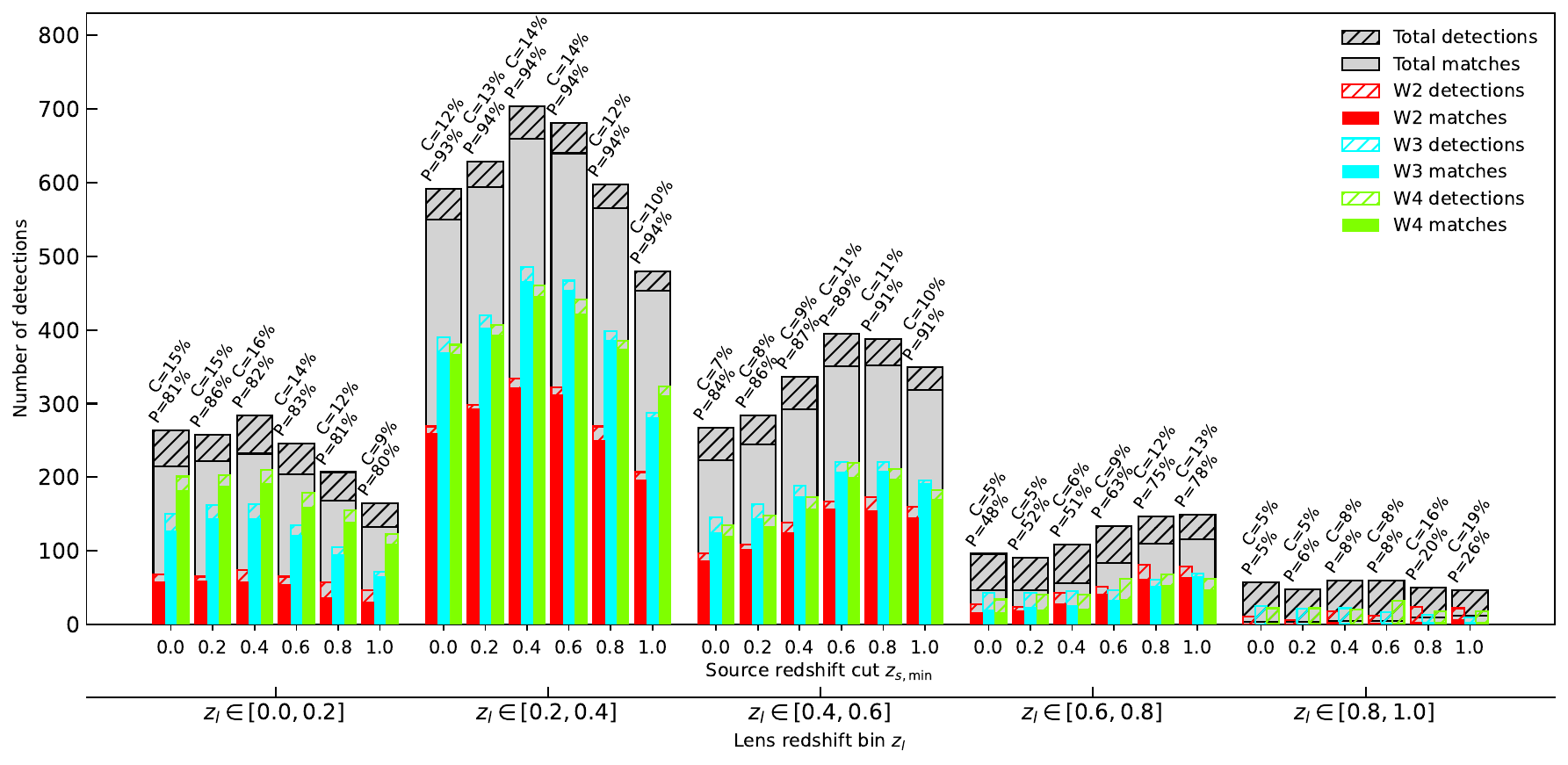}
    \caption{Histogram of the number of detections and associations on NFW mock convergence maps, produced with bins of lens redshift and source redshift. The full bars show the detections matched to the haloes input from the catalogue, while the hashed bars show the share of detections that were not matched, i.e., spurious detections. The colours show the contributions of each wavelet scale (red: W2, cyan: W3, green: W4) to the total detections/associations (grey). On top of each bar, we display purity (P) and completeness (C). The detection threshold was set on W2 as $4.6\sigma$.}
    \label{fig:nfw_hist_dets}
\end{figure*}

Using the semi-analytical NFW mocks described in Sect.~\ref{sec:data_set}, we applied our multi-scale detection algorithm to mock convergence maps generated from lens redshift slices. Specifically, we only included the lensing signal from lenses (that is, clusters) in a given redshift range, to study the idealised detectability of such lenses as a function of their redshift, while minimising the impact of line-of-sight halo superpositions. We divided the lens population into bins of width $\Delta z_l = 0.2$ between redshift 0 and 1; that is, $z_l \in [[0, 0.2], [0.2, 0.4], \ldots, [0.8, 1.0]]$. In addition, to assess the effect of dilution by foreground sources, the lensing signal was computed considering different source redshift distributions. Thus, we generated convergence maps for each lens redshift bin using different minimum source redshift cuts, $z_{s,\mathrm{min}} \in [0, 0.2, 0.4, \ldots, 1.0]$, while keeping the source redshift distribution fixed above the chosen cut. 

Figure~\ref{fig:nfw_hist_dets} shows how the detection performance evolves under these idealised conditions. Each grey bar represents the number of detections for a given lens-redshift-slice and source-redshift-cut pair, and results from the combined effect of the lens redshift distribution and source redshift distribution (via the statistical effects of dilution by foreground sources and evolution of source density with $z_{s,\mathrm{min}}$, as well as the geometrical effect of the lensing kernel). We note the following:
\begin{itemize}
    \item Fields populated with haloes with $z_l \in [0.2, 0.4]$ yield more detections, as this redshift slice  corresponds to the peak of our halo catalogue redshift distribution (see Fig.~\ref{fig:fields}).
    \item For any given lens redshift slice, fields with $z_{s,\mathrm{min}}$ just behind lens redshift $z_l$ lead to the highest numbers of detections by reducing the dilution effect.
    \item With increasing $z_{s,\mathrm{min}}$ values, the lensing kernel peaks at a higher lens redshift, but also results in a noisier convergence, as the source density decreases.
    \item Inspecting the detections per wavelet scale (coloured histograms), smaller scales become progressively more efficient with increasing $z_l$, as lenses will also tend to subtend smaller solid angles on the sky.
    \item The absolute number of spurious detections (hashed regions of the histogram) is stable throughout the different redshift bins and detection scales, demonstrating the effectiveness of our thresholding method.
\end{itemize}

\subsubsection{Comparison of source redshift binning configurations}

\begin{figure*}[h!]

    \centering
        
    \includegraphics[width=\linewidth, trim={0cm 0cm 0cm 0cm},clip]{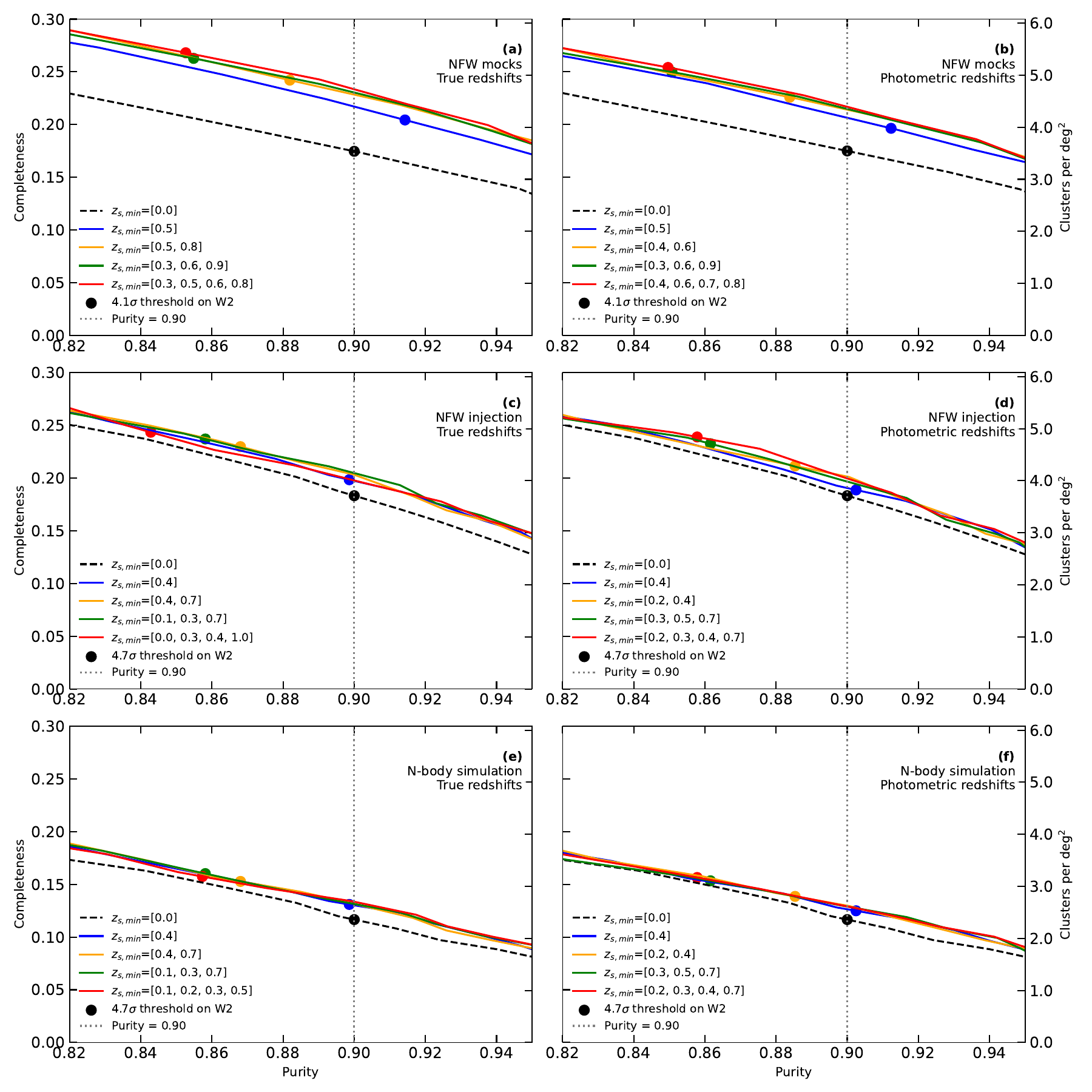}
    
    \caption{Best performing Purity-Completeness curves (ranked by AUC) for 1, 2, 3, and 4 source redshift bins (respectively blue, yellow, green, and red). Results of the integrated approach (i.e., with $z_{s,\mathrm{min}}=0$) are shown in dashed black. Full circles show the position on each corresponding curve of a detection threshold set to match 90\% purity in the integrated case.
    Left column: using the true source redshift distribution. Right column: using source redshift with {\it Euclid}-like photometric errors.
    Top row, (a) and (b): detections on NFW mock convergence maps (LSS-free, shape noise only). 
    Middle row, (c) and (d) detections of NFW mock clusters injected in the N-body simulation fields.
    Bottom row, (e) and (f): detections on the N-body simulation fields.}
    \label{fig:CvsP_best}

\end{figure*}

We will now study the impact of the tomographic approach based on the $z_{s,\mathrm{min}}$-cut on our idealised LSS-free NFW mocks. We populated matter fields by integrating the entire lens population ($z_l$ between 0 and 1, instead of the redshift slicing presented above) and produced the convergence maps by using the $z_{s,\mathrm{min}}$-cut source redshift binning ($z_{s,\mathrm{min}}$ values in $[0, 0.1, \ldots, 1]$).  
We tested four binning strategies, using 1, 2, 3, and 4 source redshift bins. In the single-bin case, we considered only one single redshift bin, but we built different configurations using all minimum redshift cuts $z_{s,\mathrm{min}}$, and we compared the number of detections. For the multi-bin cases (2, 3, and 4 bins), we explored all possible combinations of $z_{s,\mathrm{min}}$-cuts to assess how the tomographic slicing affected detection performance. This way, we aimed at empirically unveiling which binning yields the best detection performance. By varying the detection threshold, we constructed -- for each binning configuration -- the purity-completeness curves shown in Fig.~\ref{fig:CvsP_nfw_mocks_all}. The curves were then ranked as a function of their AUC in the purity range [0.85, 0.95]. The best-performing combinations are shown in the top panel of Fig.~\ref{fig:CvsP_best} for the true (zero uncertainty) source redshifts (left) and {\it Euclid}-like photometric redshifts (right) cases. 

In both cases, we observe a substantial improvement compared to the `integrated' approach (dashed black curve), in which no $z_{s,\mathrm{min}}$ cut is applied, irrespective of the number of bins. Using the best-performing single-bin analysis, $z_{s,{\mathrm{min}}}=0.5$, increases the completeness at 90\% purity by 24\% (true redshifts) and 18\% (photometric redshifts) relative to the integrated approach ($z_{s,\mathrm{min}}=0$).

When omitting photometric redshift errors, combining multiple redshift bins, that is, using multiple $z_{s,\mathrm{min}}$ cuts, leads to further improvement. Compared to the best-performing single bin case ($z_{s,{\mathrm{min}}}=0.5$), using 2, 3, or 4 bins increases the completeness at 90\% purity by 5.2\%, 6.2\%, and 7.6\%, respectively. However, accounting for realistic {\it Euclid}-like photo-$z$ reduces the multi-bin efficiency, and the completeness increase at 90\% purity compared to the best single bin method yields the values 3.7\%, 4.1\%, and 5.1\%. In both the true-$z$ and photometric-$z$ cases, the performance gain brought by combining multiple source redshift bins is small, but consistently positive across the full completeness-purity range.

\subsection{NFW mocks injected in the N-body simulated field}

To assess the impact of LSS on the detection performance, we next injected spherically symmetric NFW clusters directly into the DEMNUni-Cov realisations. This procedure was undertaken one cluster at a time, after which the wavelet detection algorithm was run on the field to assess whether the injected cluster as detected or not.  We repeated the analysis described above,  
measuring purity and completeness using all source redshift binning configurations, both with and without photometric errors (see Fig.~\ref{fig:CvsP_nfw_inj_all}). The completeness-purity curves of the best-performing source redshift binning (ranked by AUC), for each binning strategy (1, 2, 3, and 4 bins), are displayed in the middle row of Fig.~\ref{fig:CvsP_best} (true source redshifts on the left and {\it Euclid}-like photometric redshifts on the right). 

We observe that any improvement compared to the integrated approach is less pronounced as compared to the results on the LSS-free mocks, although we still note a systematic improvement: the best-performing single-bin analysis ($z_{s,{\mathrm{min}}}=0.4$) increases the completeness at 90\% purity relative to the integrated approach by 8\% (true redshifts) and 4\% (photometric redshifts). Additional  benefits from multi-bin strategies are more marginal and are not consistently positive for all values of purity.

\subsection{N-body simulations}

Finally, we ran our detection algorithm on the realisations of the N-body DEMNUni-Cov simulation, and measured completeness and purity by matching the detections to the haloes of the halo-finder catalogue, using the methodology presented in Sect.~\ref{sec:methods:matching}. We again produced purity-completeness curves for all possible combinations in the 1-, 2-, 3-, and 4-bin cases for both the true and photometric source redshifts (see Fig.~\ref{fig:CvsP_challenge_all}. Once again, we selected the best-performing curves for each case (ranked by AUC) and present them in the bottom panels of Fig.~\ref{fig:CvsP_best}, with both the true source redshifts (left) and photometric redshifts (right). 

Compared to the case where idealised NFW haloes were injected into the N-body fields, the completeness decreases by $\sim 30\%$ at fixed purity ($P=90\%$), independent of the adopted source redshift binning strategy. This reduction is expected, as realistic haloes exhibit a wider range of morphologies. However, the main conclusion drawn from the injection tests remains unchanged: the signal from the LSS nullifies any potential gain from the multi-bin strategy, as it does not yield better results than a single bin with an adequate $z_{s,\mathrm{min}}$ cut. The latter configuration still provides an improvement compared to the integrated approach, although the gain is weaker when considering photometric redshift errors (12\% with true redshifts, 8\% with photometric redshifts).

\section{Discussion}
\label{sec:discussion}

\subsection{Performance of the source redshift tomography}

In a single source redshift bin scenario, we find, in agreement with previous work \citep{Hamana_2020, Trobbiani_2025}, that imposing a lower bound to the source redshift distribution at intermediate redshift (between $z=0.3$ and $z=0.6$) yields significant improvements in purity and completeness compared to methods integrating the full source population from $z=0$. Using synthetic data of increasing complexity, we showed that both the introduction of photometric errors in source redshift measurements and the signal from LSS reduce the efficiency of the $z_{s, \mathrm{min}}$-cut method. Despite these two effects, a single bin with $z_{s, \mathrm{min}}=0.4$ still increases the completeness by 8\%  at a fixed 90\% purity on the Euclidised DEMNUni-Cov fields. However, we do not find further improvements in completeness (at fixed purity) by combining multiple bins with different $z_{s, \mathrm{min}}$-cuts.

Focusing on the coloured markers in Fig.~\ref{fig:CvsP_best}, corresponding to a fixed detection threshold, the impact of multi-bin strategies becomes clearer. Combining detections from several source redshift bins systematically shifts the markers 
    towards higher completeness,
    but simultaneously towards lower purity.
Indeed, although the source redshift bins overlap over most of the redshift range (Fig.~\ref{fig:nz_wz}) -- resulting in strongly correlated noise between bins -- the galaxies that belong exclusively to a given bin generate an independent realisation of shape noise. 

As a consequence 
    spurious detections in each bin are not identical, and by
    merging detections across bins (by taking their union), these distinct spurious peaks accumulate.
This accumulation lowers the recovered purity for a given detection threshold and substantially mitigates the potential gains provided by the multi-bin tomographic analysis. Ignoring this in a real survey, where purity cannot directly be measured, would lead to an underestimation of the contamination by spurious detections and ultimately to an overly-optimistic assessment of the benefits of multi-bin combinations. On the DEMNUni-Cov simulation fields, comparing the best performing four-bins analysis to the best performing single-bin -- at fixed threshold, that is, comparing the red and blue markers in the bottom panels of Fig.~\ref{fig:CvsP_best} -- yields a completeness increase of
    23\% /
    25\% for true redshifts and photometric redshifts, respectively.
In the same way, comparing the best multi-bin with the integrated method at a fixed threshold (red vs black markers) yields a completeness increase of:
    37\% /
    34\% for true redshifts and photometric redshifts, respectively.

As demonstrated earlier, a fairer completeness comparison at fixed purity does not show any significant improvement owing to multi-bin strategies, illustrating how misleading threshold-based comparisons can be when purity is not controlled for. We emphasise that this addition of fake detections through tomographic multi-bin merging is the dominant effect, dampening the potential improvements of multi-bin strategies. Indeed, even in our most idealised scenario (NFW mocks on Gaussian noise, free from LSS contamination and using true source redshifts), we show that a comparison at fixed threshold would suggest a completeness increase of 31\% between the best four-bins and the best single bin, far larger than the improvements of 8\% reported when comparing completeness at a fixed purity. 

\begin{figure}[]
    \centering
      \includegraphics[width=\linewidth]{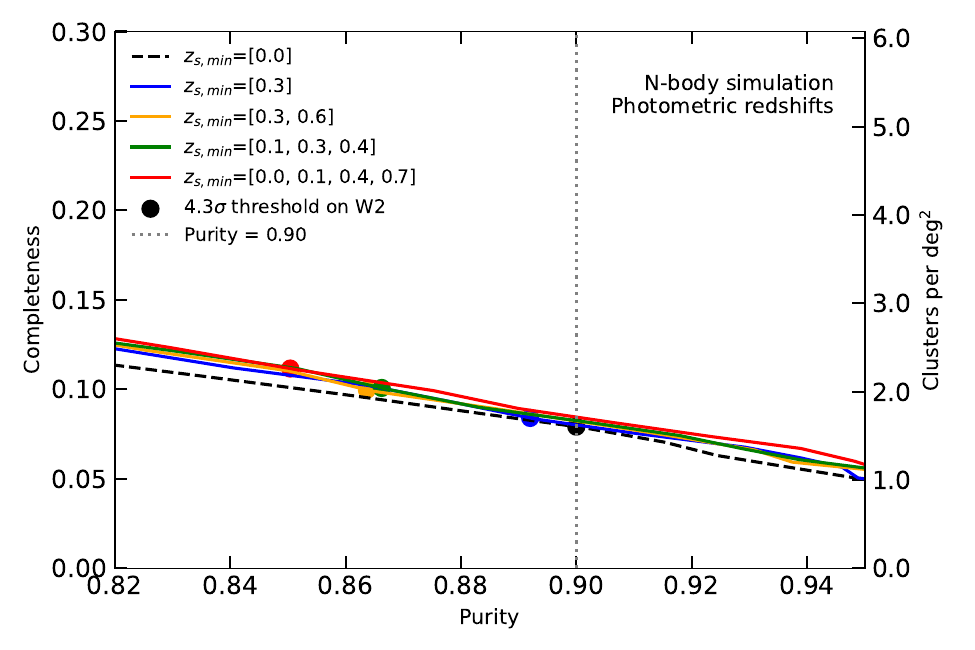}
    \caption{Tomographic detection performance for a single scale analysis (W3). Best performing Purity-Completeness curves (ranked by AUC) for 1, 2, 3, and 4 source redshift bins (respectively blue, yellow, green, and red). Results of the integrated approach (i.e., with $z_{s,\mathrm{min}}=0$) are shown in dashed black. Full circles show the position on each corresponding curve of a detection threshold set to match 90\% purity in the integrated case.
    }

    \label{fig:CvsP_best_W3}
\end{figure}

To highlight that this behaviour is not exclusive to our multi-scale approach, we also conducted our tomographic analysis on the Euclidised N-body convergence fields in a single scale mode, using the wavelet scale W3 alone. The completeness-purity curves for each bin combination (1, 2, 3, and 4 bins strategies) are shown in Appendix~\ref{fig:CvsP_challenge_all_w3}, and the best performing combinations are shown in Fig.~\ref{fig:CvsP_best_W3}. In agreement with our results using multi-scale detections, we show that multi-bin strategies do not bring significant improvement. Surveys with greater depth, accessing higher source redshifts and higher number densities, would permit finer source redshift binning and consequently narrower lensing kernels. Such conditions potentially offer better prospects for the application of tomographic strategies such as those discussed in this paper.

\subsection{Blending and association method}

\begin{table}[]
\begin{tabular}{lccc}
\hline
Scale & Single matches & Multiple matches & Unmatched \\ \hline
W2    & 0.76         & 0.11             & 0.13      \\
W3    & 0.68         & 0.23             & 0.09      \\
W4    & 0.44         & 0.50             & 0.06      \\
W234  & 0.60         & 0.30             & 0.10      \\ \hline
\end{tabular}
\caption{Fraction of single matched, multiple matched, and unmatched detections over the total number of detections, using the multi-scale formalism, or focusing on only one scale (W2, W3, or W4). Detections were performed with a global purity of 90\%, using a single redshift bin with $z_{s,\mathrm{min}}$ = 0.4, accounting for photo-z errors on the DEMNUni-Cov realisations.}
\label{table:blending}
\end{table}

\begin{figure*}[htbp] 
    \centering 
    \begin{subfigure}[b]{0.49\textwidth} 
        \centering 
        \includegraphics[width=\linewidth]{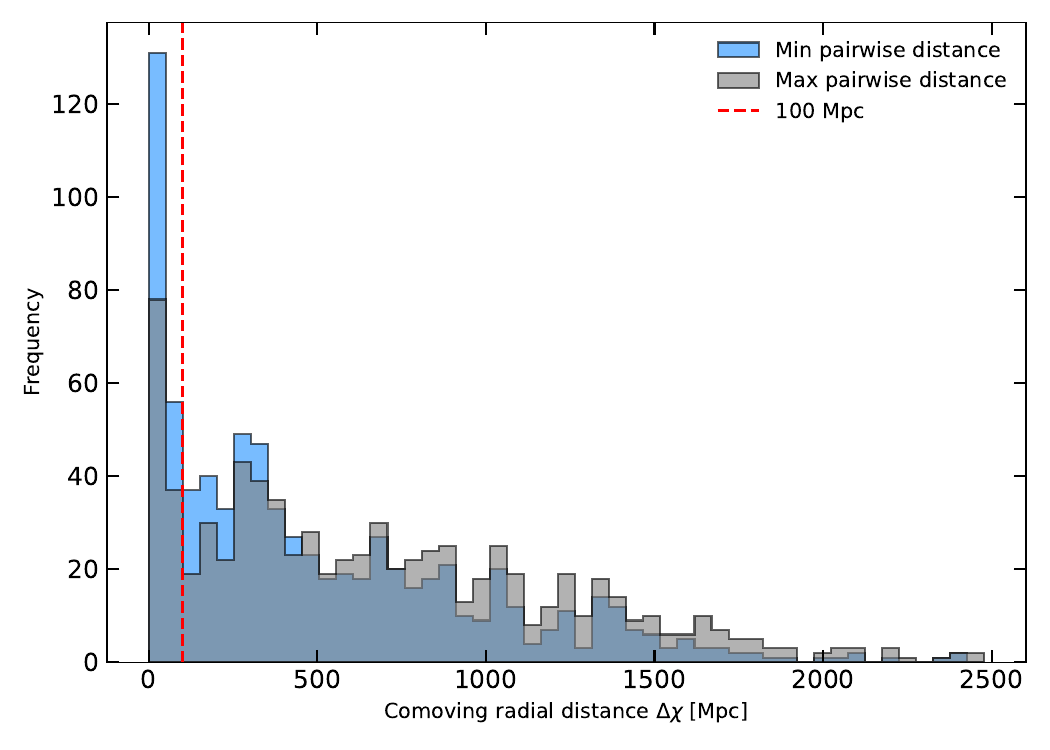} 
    \end{subfigure} 
    \hfill 
    \begin{subfigure}[b]{0.49\textwidth} 
        \centering 
        \includegraphics[width=\linewidth]{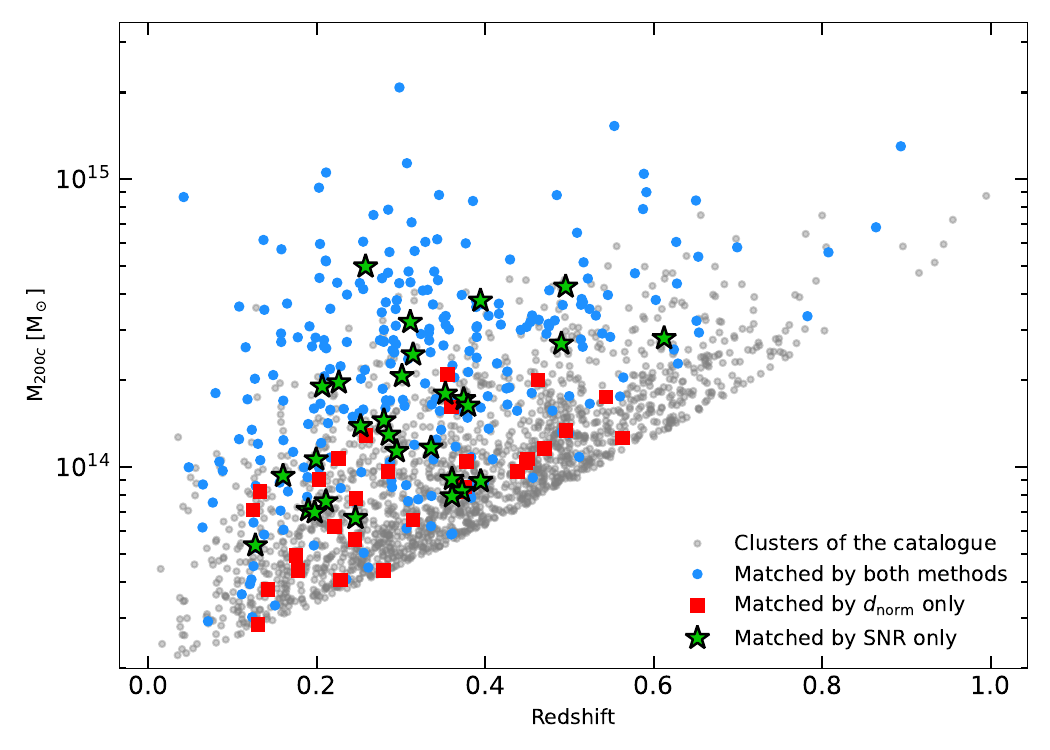} 
    \end{subfigure}

    \caption{
    Diagnostic plots for detections with multiple matching halo candidates (blended detections),
    resulting from a detection threshold of 4.7$\sigma$ on W2 and a single source redshift bin with
    $z_{s,\mathrm{min}}=0.4$, including photometric redshift uncertainties, on the DEMNUni-Cov
    realisations.
    Left: Distribution of comoving pairwise separations between clusters potentially associated with a blended detection. As blended detections can be matched to more than two clusters, closest-pair separation shown in blue, furthest-pair separation shown in grey.
    Right: Location in the mass–redshift plane of matched detections using two different discrimination criteria for blended detections: theoretical S/N (green stars) and $d_{\mathrm{norm}}$ (red squares). Common matches are shown
    in blue. The full halo catalogue considered for matching is shown in grey. Results from only one simulation field are shown.
    }
    \label{fig:blending}
\end{figure*}

In some cases, a single detection can be associated with multiple haloes. We refer to such detections as blended or multiple-matched detections, and they represent a non-negligible fraction of the total sample. In Table~\ref{table:blending}, we show the fraction of single-matched, multiple-matched, and unmatched detections relative to the total number of detections. For this table, we used the detection catalogue obtained with a single source redshift bin (with $z_{s,\mathrm{min}} = 0.4$) on the DEMNUni-Cov realisations, assuming an {\it Euclid}-like photometric source redshift distribution and photometric errors. We display the results both for the multi-scale framework and for each wavelet scale independently (W2, W3, and W4). At coarser scales, the blended fraction becomes higher, as these scales are sensitive to larger structures and can associate -- within the MMD formalism -- haloes that are more widely separated on the sky. For the largest scale, W4, the fraction of multiple matches is comparable to that of its single-matched detections. In the multi-scale case, at 90\% purity, 30\% of the total detections have multiple matching candidates. 

In this specific configuration, we investigate in Fig.~\ref{fig:blending} (left panel) the distribution of the radial distance between pairs of matching candidates for a given blended detection. We measured the comoving radial distance $\Delta\chi$ between each pair of matching candidates for each blended detection, and display the comoving radial distances between both the radially closest (blue) and most separated (grey) pairs. In both cases, we observe a bimodal distribution with a strong peak at $\Delta\chi$ values under 100~Mpc. Such cases highlight blended detections where the pair of clusters is correlated by the cosmic web. However, the number of pairs with $\Delta\chi >100\mathrm{Mpc}$ significantly outweighs these correlated clusters, showing that most blended detections come from line-of-sight alignments of uncorrelated structures.  

We compared the two ways of associating a blended detection to a unique cluster, as presented in Sect. \ref{sec:methods:matching}. Averaging over all realisations, we found the fraction of matched detections associated with a different cluster with both methods to be 10\%. In Fig.~\ref{fig:blending} (right panel), we show, for a single realisation, the positions in the mass-redshift plane of the clusters associated with each method. By construction, the clusters associated with the S/N criterion (green stars) lie preferentially toward the higher mass regions of the diagram than those associated with the $d_{\mathrm{norm}}$ criterion (red squares). The high-mass regime is where the sample is expected to be most complete, making these clusters more likely to be real contributors to the signal. Nonetheless, designating one cluster as the sole contributor to the signal in the case of blended detections may lead to biased (overestimated) mass estimates and/or an inaccurate determination of the selection function. These effects must be tackled for further cosmological analysis of WL selected samples. Additionally, further work should focus on the matching of the lensing detections with realistic optical/X-ray/SZ catalogues.

\section{Conclusions}
\label{sec:conclusion}

In this work, we investigated the detection performance of the $z_{s, \mathrm{min}}$-cut technique \citep{Hamana_2020}, which consists of combining peak detections arising in WL observables from different source redshift populations. We applied this redshift tomographic approach to the wavelet-based multi-scale algorithm presented in \citet{Leroy_2023}, to which we also brought improvements in the merging of the multi-scale detections, thresholding of the multi-scale mass maps, and association of the detected peaks with the haloes of the simulation catalogue. 

We evaluated the performance of the $z_{s, \mathrm{min}}$-cut method using either 1, 2, 3, or 4 tomographic bins. In each case, the source redshift distribution was binned with a step of $\Delta z =0.1$ in the interval [0, 1] such as $z_{s, \mathrm{min}} \in [0, 0.1, 0.2,..., 1]$, from which all $z_{s, \mathrm{min}}$ combinations were tried for a given number of bins. In each case, the purity-completeness properties of the resulting catalogue were used to quantify performance, by measuring the area under the curve in the purity range $P\in[0.85,0.95]$. We showed that a single bin strategy, with $z_{s, \mathrm{min}}$ between 0.3 and 0.6 -- a range in which dilution by foreground sources is mitigated, while keeping a relatively high source density -- performs as well as multi-bin methods. 

We applied our framework successively to (i) NFW haloes in LSS-free fields, (ii) NFW haloes injected into N-body simulated fields to account for LSS contamination while retaining idealised haloes, and finally (iii) to haloes self-consistently formed in the DEMNUni-Cov N-body simulation. In each case, we considered both perfect knowledge of the source redshift distribution and {\it Euclid}-like photometric redshift uncertainties. We find that both redshift uncertainties and LSS contamination reduce potential detection gains from tomographic methods.

However, the dominant limitation that suppresses the potential improvements from multi-bin combinations arises from the accumulation of independent spurious detections across source redshift bins. As a result, merging detections obtained from convergence maps constructed from different source redshift populations lowers the purity at a fixed detection threshold, rendering this approach inefficient for improving detection performance in a {\it Euclid}-like survey. To provide an additional test, we reverted our multi-scale detection framework to a single-scale mode, yielding the same conclusion. We emphasise that studies using combinations of detections from different tomographic bins should assess detection performance at fixed purity, rather than fixed S/N thresholds. In future work, investigating methods for the systematic flagging of spurious detections arising in different redshift bins could allow the merging of tomographic bin detections with a reduced purity loss, potentially making this technique more performant. 

Independent of the source redshift distribution, single bin, or multi-bin strategies, we also found that a non-negligible share of the detection result from the blending of signals from haloes close by in projection on the plane of the sky (but not necessarily at the same redshift). Consequently, these cannot be safely attributed to a single lens contribution. This effect naturally becomes stronger with larger filters. Defining a more accurate detection position for large-scale filters -- with smaller association distances -- in order to more accurately identify the main lens in a case where multiple haloes are within the detection filter scale, could mitigate this issue. However, for subsequent mass measurement and cosmological analysis, these blended detections will need to be accounted for.   


\begin{acknowledgements}
GWP acknowledges long-term support from CNES, the French space agency.
The DEMNUni-cov simulations were carried out in the framework of “The Dark Energy and Massive Neutrino Universe covariances" project, using the Tier-0 Intel OmniPath Cluster Marconi-A1 of the Centro Interuniversitario del Nord-Est per il Calcolo Elettronico (CINECA). CC acknowledges a generous CPU and storage allocation by the Italian Super-Computing Resource Allocation (ISCRA) as well as from the coordination of the “Accordo Quadro MoU per lo svolgimento di attività congiunta di ricerca Nuove frontiere in Astrofisica: HPC e Data Exploration di nuova generazione”, together with storage from INFN-CNAF and INAF-IA2.
\end{acknowledgements}

\bibliographystyle{aa} 
\bibliography{biblio} 

@ARTICLE{bohringer_reflex_2004,
       author = {{B{\"o}hringer}, H. and {Schuecker}, P. and {Guzzo}, L. and {Collins}, C.~A. and {Voges}, W. and {Cruddace}, R.~G. and {Ortiz-Gil}, A. and {Chincarini}, G. and {De Grandi}, S. and {Edge}, A.~C. and {MacGillivray}, H.~T. and {Neumann}, D.~M. and {Schindler}, S. and {Shaver}, P.},
        title = "{The ROSAT-ESO Flux Limited X-ray (REFLEX) Galaxy cluster survey. V. The cluster catalogue}",
      journal = {\aap},
         year = 2004,
        month = oct,
       volume = {425},
        pages = {367-383},
          doi = {10.1051/0004-6361:20034484},
archivePrefix = {arXiv},
       eprint = {astro-ph/0405546},
 primaryClass = {astro-ph},
       adsurl = {https://ui.adsabs.harvard.edu/abs/2004A&A...425..367B}
}

@ARTICLE{bulbul_erosita_2024,
       author = {{Bulbul}, E. and {Liu}, A. and {Kluge}, M. and {Zhang}, X. and {Sanders}, J.~S. and {Bahar}, Y.~E. and {Ghirardini}, V. and {Artis}, E. and {Seppi}, R. and {Garrel}, C. and {Ramos-Ceja}, M.~E. and {Comparat}, J. and {Balzer}, F. and {B{\"o}ckmann}, K. and {Br{\"u}ggen}, M. and {Clerc}, N. and {Dennerl}, K. and {Dolag}, K. and {Freyberg}, M. and {Grandis}, S. and {Gruen}, D. and {Kleinebreil}, F. and {Krippendorf}, S. and {Lamer}, G. and {Merloni}, A. and {Migkas}, K. and {Nandra}, K. and {Pacaud}, F. and {Predehl}, P. and {Reiprich}, T.~H. and {Schrabback}, T. and {Veronica}, A. and {Weller}, J. and {Zelmer}, S.},
        title = "{The SRG/eROSITA All-Sky Survey. The first catalog of galaxy clusters and groups in the Western Galactic Hemisphere}",
      journal = {\aap},
         year = 2024,
        month = may,
       volume = {685},
          eid = {A106},
        pages = {A106},
          doi = {10.1051/0004-6361/202348264},
archivePrefix = {arXiv},
       eprint = {2402.08452},
 primaryClass = {astro-ph.CO},
       adsurl = {https://ui.adsabs.harvard.edu/abs/2024A&A...685A.106B}
}

@ARTICLE{rykoff_redmapper_2014,
       author = {{Rykoff}, E.~S. and {Rozo}, E. and {Busha}, M.~T. and {Cunha}, C.~E. and {Finoguenov}, A. and {Evrard}, A. and {Hao}, J. and {Koester}, B.~P. and {Leauthaud}, A. and {Nord}, B. and {Pierre}, M. and {Reddick}, R. and {Sadibekova}, T. and {Sheldon}, E.~S. and {Wechsler}, R.~H.},
        title = "{redMaPPer. I. Algorithm and SDSS DR8 Catalog}",
      journal = {\apj},
         year = 2014,
        month = apr,
       volume = {785},
       number = {2},
          eid = {104},
        pages = {104},
          doi = {10.1088/0004-637X/785/2/104},
archivePrefix = {arXiv},
       eprint = {1303.3562},
 primaryClass = {astro-ph.CO},
       adsurl = {https://ui.adsabs.harvard.edu/abs/2014ApJ...785..104R}
}

@ARTICLE{oguri_camira_2018,
       author = {{Oguri}, Masamune and {Lin}, Yen-Ting and {Lin}, Sheng-Chieh and {Nishizawa}, Atsushi J. and {More}, Anupreeta and {More}, Surhud and {Hsieh}, Bau-Ching and {Medezinski}, Elinor and {Miyatake}, Hironao and {Jian}, Hung-Yu and {Lin}, Lihwai and {Takada}, Masahiro and {Okabe}, Nobuhiro and {Speagle}, Joshua S. and {Coupon}, Jean and {Leauthaud}, Alexie and {Lupton}, Robert H. and {Miyazaki}, Satoshi and {Price}, Paul A. and {Tanaka}, Masayuki and {Chiu}, I.-Non and {Komiyama}, Yutaka and {Okura}, Yuki and {Tanaka}, Manobu M. and {Usuda}, Tomonori},
        title = "{An optically-selected cluster catalog at redshift 0.1 < z < 1.1 from the Hyper Suprime-Cam Subaru Strategic Program S16A data}",
      journal = {\pasj},
         year = 2018,
        month = jan,
       volume = {70},
          eid = {S20},
        pages = {S20},
          doi = {10.1093/pasj/psx042},
archivePrefix = {arXiv},
       eprint = {1701.00818},
 primaryClass = {astro-ph.CO},
       adsurl = {https://ui.adsabs.harvard.edu/abs/2018PASJ...70S..20O}
}

@ARTICLE{bleem_spt_2015,
       author = {{Bleem}, L.~E. and {Stalder}, B. and {de Haan}, T. and {Aird}, K.~A. and {Allen}, S.~W. and {Applegate}, D.~E. and {Ashby}, M.~L.~N. and {Bautz}, M. and {Bayliss}, M. and {Benson}, B.~A. and {Bocquet}, S. and {Brodwin}, M. and {Carlstrom}, J.~E. and {Chang}, C.~L. and {Chiu}, I. and {Cho}, H.~M. and {Clocchiatti}, A. and {Crawford}, T.~M. and {Crites}, A.~T. and {Desai}, S. and {Dietrich}, J.~P. and {Dobbs}, M.~A. and {Foley}, R.~J. and {Forman}, W.~R. and {George}, E.~M. and {Gladders}, M.~D. and {Gonzalez}, A.~H. and {Halverson}, N.~W. and {Hennig}, C. and {Hoekstra}, H. and {Holder}, G.~P. and {Holzapfel}, W.~L. and {Hrubes}, J.~D. and {Jones}, C. and {Keisler}, R. and {Knox}, L. and {Lee}, A.~T. and {Leitch}, E.~M. and {Liu}, J. and {Lueker}, M. and {Luong-Van}, D. and {Mantz}, A. and {Marrone}, D.~P. and {McDonald}, M. and {McMahon}, J.~J. and {Meyer}, S.~S. and {Mocanu}, L. and {Mohr}, J.~J. and {Murray}, S.~S. and {Padin}, S. and {Pryke}, C. and {Reichardt}, C.~L. and {Rest}, A. and {Ruel}, J. and {Ruhl}, J.~E. and {Saliwanchik}, B.~R. and {Saro}, A. and {Sayre}, J.~T. and {Schaffer}, K.~K. and {Schrabback}, T. and {Shirokoff}, E. and {Song}, J. and {Spieler}, H.~G. and {Stanford}, S.~A. and {Staniszewski}, Z. and {Stark}, A.~A. and {Story}, K.~T. and {Stubbs}, C.~W. and {Vanderlinde}, K. and {Vieira}, J.~D. and {Vikhlinin}, A. and {Williamson}, R. and {Zahn}, O. and {Zenteno}, A.},
        title = "{Galaxy Clusters Discovered via the Sunyaev-Zel'dovich Effect in the 2500-Square-Degree SPT-SZ Survey}",
      journal = {\apjs},
         year = 2015,
        month = feb,
       volume = {216},
       number = {2},
          eid = {27},
        pages = {27},
          doi = {10.1088/0067-0049/216/2/27},
archivePrefix = {arXiv},
       eprint = {1409.0850},
 primaryClass = {astro-ph.CO},
       adsurl = {https://ui.adsabs.harvard.edu/abs/2015ApJS..216...27B}
}

@ARTICLE{hilton_act_2021,
       author = {{Hilton}, M. and {Sif{\'o}n}, C. and {Naess}, S. and {Madhavacheril}, M. and {Oguri}, M. and {Rozo}, E. and {Rykoff}, E. and {Abbott}, T.~M.~C. and {Adhikari}, S. and {Aguena}, M. and {Aiola}, S. and {Allam}, S. and {Amodeo}, S. and {Amon}, A. and {Annis}, J. and {Ansarinejad}, B. and {Aros-Bunster}, C. and {Austermann}, J.~E. and {Avila}, S. and {Bacon}, D. and {Battaglia}, N. and {Beall}, J.~A. and {Becker}, D.~T. and {Bernstein}, G.~M. and {Bertin}, E. and {Bhandarkar}, T. and {Bhargava}, S. and {Bond}, J.~R. and {Brooks}, D. and {Burke}, D.~L. and {Calabrese}, E. and {Carrasco Kind}, M. and {Carretero}, J. and {Choi}, S.~K. and {Choi}, A. and {Conselice}, C. and {da Costa}, L.~N. and {Costanzi}, M. and {Crichton}, D. and {Crowley}, K.~T. and {D{\"u}nner}, R. and {Denison}, E.~V. and {Devlin}, M.~J. and {Dicker}, S.~R. and {Diehl}, H.~T. and {Dietrich}, J.~P. and {Doel}, P. and {Duff}, S.~M. and {Duivenvoorden}, A.~J. and {Dunkley}, J. and {Everett}, S. and {Ferraro}, S. and {Ferrero}, I. and {Fert{\'e}}, A. and {Flaugher}, B. and {Frieman}, J. and {Gallardo}, P.~A. and {Garc{\'\i}a-Bellido}, J. and {Gaztanaga}, E. and {Gerdes}, D.~W. and {Giles}, P. and {Golec}, J.~E. and {Gralla}, M.~B. and {Grandis}, S. and {Gruen}, D. and {Gruendl}, R.~A. and {Gschwend}, J. and {Gutierrez}, G. and {Han}, D. and {Hartley}, W.~G. and {Hasselfield}, M. and {Hill}, J.~C. and {Hilton}, G.~C. and {Hincks}, A.~D. and {Hinton}, S.~R. and {Ho}, S.-P.~P. and {Honscheid}, K. and {Hoyle}, B. and {Hubmayr}, J. and {Huffenberger}, K.~M. and {Hughes}, J.~P. and {Jaelani}, A.~T. and {Jain}, B. and {James}, D.~J. and {Jeltema}, T. and {Kent}, S. and {Knowles}, K. and {Koopman}, B.~J. and {Kuehn}, K. and {Lahav}, O. and {Lima}, M. and {Lin}, Y.-T. and {Lokken}, M. and {Loubser}, S.~I. and {MacCrann}, N. and {Maia}, M.~A.~G. and {Marriage}, T.~A. and {Martin}, J. and {McMahon}, J. and {Melchior}, P. and {Menanteau}, F. and {Miquel}, R. and {Miyatake}, H. and {Moodley}, K. and {Morgan}, R. and {Mroczkowski}, T. and {Nati}, F. and {Newburgh}, L.~B. and {Niemack}, M.~D. and {Nishizawa}, A.~J. and {Ogando}, R.~L.~C. and {Orlowski-Scherer}, J. and {Page}, L.~A. and {Palmese}, A. and {Partridge}, B. and {Paz-Chinch{\'o}n}, F. and {Phakathi}, P. and {Plazas}, A.~A. and {Robertson}, N.~C. and {Romer}, A.~K. and {Carnero Rosell}, A. and {Salatino}, M. and {Sanchez}, E. and {Schaan}, E. and {Schillaci}, A. and {Sehgal}, N. and {Serrano}, S. and {Shin}, T. and {Simon}, S.~M. and {Smith}, M. and {Soares-Santos}, M. and {Spergel}, D.~N. and {Staggs}, S.~T. and {Storer}, E.~R. and {Suchyta}, E. and {Swanson}, M.~E.~C. and {Tarle}, G. and {Thomas}, D. and {To}, C. and {Trac}, H. and {Ullom}, J.~N. and {Vale}, L.~R. and {Van Lanen}, J. and {Vavagiakis}, E.~M. and {De Vicente}, J. and {Wilkinson}, R.~D. and {Wollack}, E.~J. and {Xu}, Z. and {Zhang}, Y.},
        title = "{The Atacama Cosmology Telescope: A Catalog of >4000 Sunyaev-Zel{\textquoteright}dovich Galaxy Clusters}",
      journal = {\apjs},
         year = 2021,
        month = mar,
       volume = {253},
       number = {1},
          eid = {3},
        pages = {3},
          doi = {10.3847/1538-4365/abd023},
archivePrefix = {arXiv},
       eprint = {2009.11043},
 primaryClass = {astro-ph.CO},
       adsurl = {https://ui.adsabs.harvard.edu/abs/2021ApJS..253....3H}
}

@ARTICLE{giocoli17,
   author = {{Giocoli}, C. and {Di Meo}, S. and {Meneghetti}, M. and {Jullo}, E. and 
	{de la Torre}, S. and {Moscardini}, L. and {Baldi}, M. and {Mazzotta}, P. and 
	{Metcalf}, R.~B.},
    title = "{Fast weak-lensing simulations with halo model}",
  journal = {\mnras},
archivePrefix = "arXiv",
   eprint = {1701.02739},
     year = 2017,
    month = sep,
   volume = 470,
    pages = {3574-3590},
      doi = {10.1093/mnras/stx1399},
   adsurl = {http://adsabs.harvard.edu/abs/2017MNRAS.470.3574G}
}

@ARTICLE{planckxx,
   author = {{Planck Collaboration XX}},
    title = "{Planck 2013 results. XX. Cosmology from Sunyaev-Zeldovich cluster counts}",
  journal = {\aap},
archivePrefix = "arXiv",
   eprint = {1303.5080},
     year = 2014,
    month = nov,
   volume = 571,
      eid = {A20},
    pages = {A20},
      doi = {10.1051/0004-6361/201321521},
   adsurl = {http://adsabs.harvard.edu/abs/2014A%26A...571A..20P}
}

@ARTICLE{planckxxiv,
   author = {{Planck Collaboration XXIV}},
    title = "{Planck 2015 results. XXIV. Cosmology from Sunyaev-Zeldovich cluster counts}",
  journal = {\aap},
archivePrefix = "arXiv",
   eprint = {1502.01597},
     year = 2016,
    month = sep,
   volume = 594,
      eid = {A24},
    pages = {A24},
      doi = {10.1051/0004-6361/201525833},
   adsurl = {http://adsabs.harvard.edu/abs/2016A%26A...594A..24P}
}

@article{bardeen86,
	Adsurl = {http://adsabs.harvard.edu/abs/1986ApJ...304...15B},
	Author = {{Bardeen}, J.~M. and {Bond}, J.~R. and {Kaiser}, N. and {Szalay}, A.~S.},
	Doi = {10.1086/164143},
	Journal = {\apj},
	Month = may,
	Pages = {15-61},
	Title = {{The statistics of peaks of Gaussian random fields}},
	Volume = 304,
	Year = 1986}

@article{tinker08,
	Adsurl = {http://adsabs.harvard.edu/abs/2008ApJ...688..709T},
	Archiveprefix = {arXiv},
	Author = {{Tinker}, J. and {Kravtsov}, A.~V. and {Klypin}, A. and {Abazajian}, K. and {Warren}, M. and {Yepes}, G. and {Gottl{\"o}ber}, S. and {Holz}, D.~E.},
	Doi = {10.1086/591439},
	Eprint = {0803.2706},
	Journal = {\apj},
	Month = dec,
	Pages = {709-728},
	Title = {{Toward a Halo Mass Function for Precision Cosmology: The Limits of Universality}},
	Volume = 688,
	Year = 2008}

@ARTICLE{despali16,
   author = {{Despali}, G. and {Giocoli}, C. and {Angulo}, R.~E. and {Tormen}, G. and
	{Sheth}, R.~K. and {Baso}, G. and {Moscardini}, L.},
    title = "{The universality of the virial halo mass function and models for non-universality of other halo definitions}",
  journal = {\mnras},
archivePrefix = "arXiv",
   eprint = {1507.05627},
     year = 2016,
    month = mar,
   volume = 456,
    pages = {2486-2504},
      doi = {10.1093/mnras/stv2842},
   adsurl = {http://adsabs.harvard.edu/abs/2016MNRAS.456.2486D}
}

@ARTICLE{EuclidSkyOverview,
author = {{Euclid Collaboration: Mellier}, Y. and {Abdurro'uf} and {Acevedo~Barroso}, J.A. and others},
	title = {Euclid - I. Overview of the Euclid mission},
	DOI= "10.1051/0004-6361/202450810",
	url= "https://doi.org/10.1051/0004-6361/202450810",
	journal = {A\&A},
	year = 2025,
	volume = 697,
	pages = "A1",
}

@INPROCEEDINGS{Roman_2024,
       author = {{Schlieder}, Joshua E. and {Barclay}, Thomas and {Barnes}, Amethyst and {Bray}, Evan and {Choi}, Ami and {Cromey}, Benjamin and {Delker}, Thomas and {Finch}, Timothy and {Frater}, Eric H. and {Hill}, Robert J. and {Kruk}, Jeffrey and {Lasco}, Jeffrey and {Louie}, Dana R. and {Malhotra}, Sangeeta and {McEnery}, Julie E. and {Mosby}, Gregory and {Paine}, Jennie and {Perkins}, Jeremy S. and {Rauscher}, Bernard J. and {Rhoads}, James E. and {Rizzo}, Maxime and {Sabatke}, Derek and {Schweickart}, Rusty and {Shukis}, Diana and {Switzer}, Eric R. and {Wollack}, Edward J. and {Zellem}, Robert T. and {Zimmerman}, Neil T.},
        title = "{Survey science with the Nancy Grace Roman Space Telescope Wide Field Instrument}",
    booktitle = {Space Telescopes and Instrumentation 2024: Optical, Infrared, and Millimeter Wave},
         year = 2024,
       editor = {{Coyle}, Laura E. and {Matsuura}, Shuji and {Perrin}, Marshall D.},
       series = {Society of Photo-Optical Instrumentation Engineers (SPIE) Conference Series},
       volume = {13092},
        month = aug,
          eid = {130920S},
        pages = {130920S},
          doi = {10.1117/12.3020622},
       adsurl = {https://ui.adsabs.harvard.edu/abs/2024SPIE13092E..0SS}
}

@ARTICLE{LSST_2019,
       author = {{Ivezi{\'c}}, {\v{Z}}eljko and {Kahn}, Steven M. and {Tyson}, J. Anthony and {Abel}, Bob and {Acosta}, Emily and {Allsman}, Robyn and {Alonso}, David and {AlSayyad}, Yusra and {Anderson}, Scott F. and {Andrew}, John and {Angel}, James Roger P. and {Angeli}, George Z. and {Ansari}, Reza and {Antilogus}, Pierre and {Araujo}, Constanza and {Armstrong}, Robert and {Arndt}, Kirk T. and {Astier}, Pierre and {Aubourg}, {\'E}ric and {Auza}, Nicole and {Axelrod}, Tim S. and {Bard}, Deborah J. and {Barr}, Jeff D. and {Barrau}, Aurelian and {Bartlett}, James G. and {Bauer}, Amanda E. and {Bauman}, Brian J. and {Baumont}, Sylvain and {Bechtol}, Ellen and {Bechtol}, Keith and {Becker}, Andrew C. and {Becla}, Jacek and {Beldica}, Cristina and {Bellavia}, Steve and {Bianco}, Federica B. and {Biswas}, Rahul and {Blanc}, Guillaume and {Blazek}, Jonathan and {Blandford}, Roger D. and {Bloom}, Josh S. and {Bogart}, Joanne and {Bond}, Tim W. and {Booth}, Michael T. and {Borgland}, Anders W. and {Borne}, Kirk and {Bosch}, James F. and {Boutigny}, Dominique and {Brackett}, Craig A. and {Bradshaw}, Andrew and {Brandt}, William Nielsen and {Brown}, Michael E. and {Bullock}, James S. and {Burchat}, Patricia and {Burke}, David L. and {Cagnoli}, Gianpietro and {Calabrese}, Daniel and {Callahan}, Shawn and {Callen}, Alice L. and {Carlin}, Jeffrey L. and {Carlson}, Erin L. and {Chandrasekharan}, Srinivasan and {Charles-Emerson}, Glenaver and {Chesley}, Steve and {Cheu}, Elliott C. and {Chiang}, Hsin-Fang and {Chiang}, James and {Chirino}, Carol and {Chow}, Derek and {Ciardi}, David R. and {Claver}, Charles F. and {Cohen-Tanugi}, Johann and {Cockrum}, Joseph J. and {Coles}, Rebecca and {Connolly}, Andrew J. and {Cook}, Kem H. and {Cooray}, Asantha and {Covey}, Kevin R. and {Cribbs}, Chris and {Cui}, Wei and {Cutri}, Roc and {Daly}, Philip N. and {Daniel}, Scott F. and {Daruich}, Felipe and {Daubard}, Guillaume and {Daues}, Greg and {Dawson}, William and {Delgado}, Francisco and {Dellapenna}, Alfred and {de Peyster}, Robert and {de Val-Borro}, Miguel and {Digel}, Seth W. and {Doherty}, Peter and {Dubois}, Richard and {Dubois-Felsmann}, Gregory P. and {Durech}, Josef and {Economou}, Frossie and {Eifler}, Tim and {Eracleous}, Michael and {Emmons}, Benjamin L. and {Fausti Neto}, Angelo and {Ferguson}, Henry and {Figueroa}, Enrique and {Fisher-Levine}, Merlin and {Focke}, Warren and {Foss}, Michael D. and {Frank}, James and {Freemon}, Michael D. and {Gangler}, Emmanuel and {Gawiser}, Eric and {Geary}, John C. and {Gee}, Perry and {Geha}, Marla and {Gessner}, Charles J.~B. and {Gibson}, Robert R. and {Gilmore}, D. Kirk and {Glanzman}, Thomas and {Glick}, William and {Goldina}, Tatiana and {Goldstein}, Daniel A. and {Goodenow}, Iain and {Graham}, Melissa L. and {Gressler}, William J. and {Gris}, Philippe and {Guy}, Leanne P. and {Guyonnet}, Augustin and {Haller}, Gunther and {Harris}, Ron and {Hascall}, Patrick A. and {Haupt}, Justine and {Hernandez}, Fabio and {Herrmann}, Sven and {Hileman}, Edward and {Hoblitt}, Joshua and {Hodgson}, John A. and {Hogan}, Craig and {Howard}, James D. and {Huang}, Dajun and {Huffer}, Michael E. and {Ingraham}, Patrick and {Innes}, Walter R. and {Jacoby}, Suzanne H. and {Jain}, Bhuvnesh and {Jammes}, Fabrice and {Jee}, M. James and {Jenness}, Tim and {Jernigan}, Garrett and {Jevremovi{\'c}}, Darko and {Johns}, Kenneth and {Johnson}, Anthony S. and {Johnson}, Margaret W.~G. and {Jones}, R. Lynne and {Juramy-Gilles}, Claire and {Juri{\'c}}, Mario and {Kalirai}, Jason S. and {Kallivayalil}, Nitya J. and {Kalmbach}, Bryce and {Kantor}, Jeffrey P. and {Karst}, Pierre and {Kasliwal}, Mansi M. and {Kelly}, Heather and {Kessler}, Richard and {Kinnison}, Veronica and {Kirkby}, David and {Knox}, Lloyd and {Kotov}, Ivan V. and {Krabbendam}, Victor L. and {Krughoff}, K. Simon and {Kub{\'a}nek}, Petr and {Kuczewski}, John and {Kulkarni}, Shri and {Ku}, John and {Kurita}, Nadine R. and {Lage}, Craig S. and {Lambert}, Ron and {Lange}, Travis and {Langton}, J. Brian and {Le Guillou}, Laurent and {Levine}, Deborah and {Liang}, Ming and {Lim}, Kian-Tat and {Lintott}, Chris J. and {Long}, Kevin E. and {Lopez}, Margaux and {Lotz}, Paul J. and {Lupton}, Robert H. and {Lust}, Nate B. and {MacArthur}, Lauren A. and {Mahabal}, Ashish and {Mandelbaum}, Rachel and {Markiewicz}, Thomas W. and {Marsh}, Darren S. and {Marshall}, Philip J. and {Marshall}, Stuart and {May}, Morgan and {McKercher}, Robert and {McQueen}, Michelle and {Meyers}, Joshua and {Migliore}, Myriam and {Miller}, Michelle and {Mills}, David J.},
        title = "{LSST: From Science Drivers to Reference Design and Anticipated Data Products}",
      journal = {\apj},
         year = 2019,
        month = mar,
       volume = {873},
       number = {2},
          eid = {111},
        pages = {111},
          doi = {10.3847/1538-4357/ab042c},
archivePrefix = {arXiv},
       eprint = {0805.2366},
 primaryClass = {astro-ph},
       adsurl = {https://ui.adsabs.harvard.edu/abs/2019ApJ...873..111I}
}

@ARTICLE{Euclid_Lesci_2024,
       author = {{Euclid Collaboration: Lesci}, G.~F. and {Sereno}, M. and {Radovich}, M. and others},
        title = "{Euclid preparation. XXXVII. Galaxy colour selections with \Euclid and ground photometry for cluster weak-lensing analyses}",
      journal = {\aap},
         year = 2024,
        month = apr,
       volume = {684},
          eid = {A139},
        pages = {A139},
          doi = {10.1051/0004-6361/202348743},
archivePrefix = {arXiv},
       eprint = {2311.16239},
 primaryClass = {astro-ph.CO},
       adsurl = {https://ui.adsabs.harvard.edu/abs/2024A&A...684A.139E}
}

@ARTICLE{Euclid_Ajani_2023,
       author = {{Euclid Collaboration: Ajani}, V. and {Baldi}, M. and {Barthelemy}, A. and others},
        title = "{\Euclid preparation. XXVIII. Forecasts for ten different higher-order weak lensing statistics}",
      journal = {\aap},
         year = 2023,
        month = jul,
       volume = {675},
          eid = {A120},
        pages = {A120},
          doi = {10.1051/0004-6361/202346017},
archivePrefix = {arXiv},
       eprint = {2301.12890},
 primaryClass = {astro-ph.CO},
       adsurl = {https://ui.adsabs.harvard.edu/abs/2023A&A...675A.120E}
}

@ARTICLE{Euclid_Martinet_2019,
       author = {{Euclid Collaboration: Martinet}, N. and {Schrabback}, T. and {Hoekstra}, H. and others},
        title = "{Euclid preparation. IV. Impact of undetected galaxies on weak-lensing shear measurements}",
      journal = {\aap},
         year = 2019,
        month = jul,
       volume = {627},
          eid = {A59},
        pages = {A59},
          doi = {10.1051/0004-6361/201935187},
archivePrefix = {arXiv},
       eprint = {1902.00044},
 primaryClass = {astro-ph.CO},
       adsurl = {https://ui.adsabs.harvard.edu/abs/2019A&A...627A..59E}
}

@article{Allen_2011,
   title={Cosmological Parameters from Observations of Galaxy Clusters},
   volume={49},
   ISSN={1545-4282},
   url={http://dx.doi.org/10.1146/annurev-astro-081710-102514},
   DOI={10.1146/annurev-astro-081710-102514},
   number={1},
   journal={Annual Review of A\&A},
   publisher={Annual Reviews},
   author={Allen, Steven W. and Evrard, August E. and Mantz, Adam B.},
   year={2011},
   month=sep, pages={409–470} }

@ARTICLE{Andreon_2012,
       author = {{Andreon}, S. and {Berg{\'e}}, J.},
        title = "{Richness-mass relation self-calibration for galaxy clusters}",
      journal = {\aap},
         year = 2012,
        month = nov,
       volume = {547},
          eid = {A117},
        pages = {A117},
          doi = {10.1051/0004-6361/201220115},
archivePrefix = {arXiv},
       eprint = {1209.5938},
 primaryClass = {astro-ph.CO},
       adsurl = {https://ui.adsabs.harvard.edu/abs/2012A&A...547A.117A}
}

@ARTICLE{Bartelmann&Schneider_2001,
       author = {{Bartelmann}, M. and {Schneider}, P.},
        title = "{Weak gravitational lensing}",
      journal = {\physrep},
         year = 2001,
        month = jan,
       volume = {340},
       number = {4-5},
        pages = {291-472},
          doi = {10.1016/S0370-1573(00)00082-X},
archivePrefix = {arXiv},
       eprint = {astro-ph/9912508},
 primaryClass = {astro-ph},
       adsurl = {https://ui.adsabs.harvard.edu/abs/2001PhR...340..291B}
}

@ARTICLE{Berge_2010,
       author = {{Berg{\'e}}, Joel and {Amara}, Adam and {R{\'e}fr{\'e}gier}, Alexandre},
        title = "{Optimal Capture of Non-Gaussianity in Weak-Lensing Surveys: Power Spectrum, Bispectrum, and Halo Counts}",
      journal = {\apj},
         year = 2010,
        month = apr,
       volume = {712},
       number = {2},
        pages = {992-1002},
          doi = {10.1088/0004-637X/712/2/992},
archivePrefix = {arXiv},
       eprint = {0909.0529},
 primaryClass = {astro-ph.CO},
       adsurl = {https://ui.adsabs.harvard.edu/abs/2010ApJ...712..992B}
}

@ARTICLE{Carbone_2016,
       author = {{Carbone}, Carmelita and {Petkova}, Margarita and {Dolag}, Klaus},
        title = "{DEMNUni: ISW, Rees-Sciama, and weak-lensing in the presence of massive neutrinos}",
      journal = {\jcap},
         year = 2016,
        month = jul,
       volume = {2016},
       number = {7},
          eid = {034},
        pages = {034},
          doi = {10.1088/1475-7516/2016/07/034},
archivePrefix = {arXiv},
       eprint = {1605.02024},
 primaryClass = {astro-ph.CO},
       adsurl = {https://ui.adsabs.harvard.edu/abs/2016JCAP...07..034C}
}

@article{Chiu_2018,
   title={Baryon content in a sample of 91 galaxy clusters selected by the South Pole Telescope at 0.2<z<1.25},
   volume={478},
   ISSN={1365-2966},
   url={http://dx.doi.org/10.1093/mnras/sty1284},
   DOI={10.1093/mnras/sty1284},
   number={3},
   journal={MNRAS},
   publisher={Oxford University Press (OUP)},
   author={Chiu, I and Mohr, J J and McDonald, M and Bocquet, S and Desai, S and Klein, M and Israel, H and Ashby, M L N and Stanford, A and Benson, B A and Brodwin, M and Abbott, T M C and Abdalla, F B and Allam, S and Annis, J and Bayliss, M and Benoit-Lévy, A and Bertin, E and Bleem, L and Brooks, D and Buckley-Geer, E and Bulbul, E and Capasso, R and Carlstrom, J E and Rosell, A Carnero and Carretero, J and Castander, F J and Cunha, C E and D’Andrea, C B and da Costa, L N and Davis, C and Diehl, H T and Dietrich, J P and Doel, P and Drlica-Wagner, A and Eifler, T F and Evrard, A E and Flaugher, B and García-Bellido, J and Garmire, G and Gaztanaga, E and Gerdes, D W and Gonzalez, A and Gruen, D and Gruendl, R A and Gschwend, J and Gupta, N and Gutierrez, G and Hlavacek-L, J and Honscheid, K and James, D J and Jeltema, T and Kraft, R and Krause, E and Kuehn, K and Kuhlmann, S and Kuropatkin, N and Lahav, O and Lima, M and Maia, M A G and Marshall, J L and Melchior, P and Menanteau, F and Miquel, R and Murray, S and Nord, B and Ogando, R L C and Plazas, A A and Rapetti, D and Reichardt, C L and Romer, A K and Roodman, A and Sanchez, E and Saro, A and Scarpine, V and Schindler, R and Schubnell, M and Sharon, K and Smith, R C and Smith, M and Soares-Santos, M and Sobreira, F and Stalder, B and Stern, C and Strazzullo, V and Suchyta, E and Swanson, M E C and Tarle, G and Vikram, V and Walker, A R and Weller, J and Zhang, Y},
   year={2018},
   month=may, pages={3072–3099} }

@ARTICLE{Chiu_2024,
       author = {{Chiu}, I. -Non and {Chen}, Kai-Feng and {Oguri}, Masamune and {Rau}, Markus M. and {Hamana}, Takashi and {Lin}, Yen-Ting and {Miyatake}, Hironao and {Miyazaki}, Satoshi and {More}, Surhud and {Sunayama}, Tomomi and {Sugiyama}, Sunao and {Takada}, Masahiro},
        title = "{Weak-Lensing Shear-Selected Galaxy Clusters from the Hyper Suprime-Cam Subaru Strategic Program: II. Cosmological Constraints from the Cluster Abundance}",
      journal = {The Open Journal of Astrophysics},
         year = 2024,
        month = oct,
       volume = {7},
          eid = {90},
        pages = {90},
          doi = {10.33232/001c.124537},
archivePrefix = {arXiv},
       eprint = {2406.11970},
 primaryClass = {astro-ph.CO},
       adsurl = {https://ui.adsabs.harvard.edu/abs/2024OJAp....7E..90C}
}

@ARTICLE{Davis_1985,
       author = {{Davis}, M. and {Efstathiou}, G. and {Frenk}, C.~S. and {White}, S.~D.~M.},
        title = "{The evolution of large-scale structure in a universe dominated by cold dark matter}",
      journal = {\apj},
         year = 1985,
        month = may,
       volume = {292},
        pages = {371-394},
          doi = {10.1086/163168},
       adsurl = {https://ui.adsabs.harvard.edu/abs/1985ApJ...292..371D}
}

@ARTICLE{Diemer_2018,
       author = {{Diemer}, Benedikt},
        title = "{COLOSSUS: A Python Toolkit for Cosmology, Large-scale Structure, and Dark Matter Halos}",
      journal = {\apjs},
         year = 2018,
        month = dec,
       volume = {239},
       number = {2},
          eid = {35},
        pages = {35},
          doi = {10.3847/1538-4365/aaee8c},
archivePrefix = {arXiv},
       eprint = {1712.04512},
 primaryClass = {astro-ph.CO},
       adsurl = {https://ui.adsabs.harvard.edu/abs/2018ApJS..239...35D}
}

@ARTICLE{Eckert_2011,
       author = {{Eckert}, D. and {Molendi}, S. and {Paltani}, S.},
        title = "{The cool-core bias in X-ray galaxy cluster samples. I. Method and application to HIFLUGCS}",
      journal = {\aap},
         year = 2011,
        month = feb,
       volume = {526},
          eid = {A79},
        pages = {A79},
          doi = {10.1051/0004-6361/201015856},
archivePrefix = {arXiv},
       eprint = {1011.3302},
 primaryClass = {astro-ph.CO},
       adsurl = {https://ui.adsabs.harvard.edu/abs/2011A&A...526A..79E}
}

@ARTICLE{Giocoli_2014,
       author = {{Giocoli}, Carlo and {Meneghetti}, Massimo and {Metcalf}, R. Benton and {Ettori}, Stefano and {Moscardini}, Lauro},
        title = "{Mass and concentration estimates from weak and strong gravitational lensing: a systematic study}",
      journal = {\mnras},
         year = 2014,
        month = may,
       volume = {440},
       number = {2},
        pages = {1899-1915},
          doi = {10.1093/mnras/stu303},
archivePrefix = {arXiv},
       eprint = {1311.1205},
 primaryClass = {astro-ph.CO},
       adsurl = {https://ui.adsabs.harvard.edu/abs/2014MNRAS.440.1899G}
}

@article{Gonzalez_2013,
   title={GALAXY CLUSTER BARYON FRACTIONS REVISITED},
   volume={778},
   ISSN={1538-4357},
   url={http://dx.doi.org/10.1088/0004-637X/778/1/14},
   DOI={10.1088/0004-637x/778/1/14},
   number={1},
   journal={ApJ},
   publisher={American Astronomical Society},
   author={Gonzalez, Anthony H. and Sivanandam, Suresh and Zabludoff, Ann I. and Zaritsky, Dennis},
   year={2013},
   month=oct, pages={14} }

@ARTICLE{Hamana_2012,
       author = {{Hamana}, Takashi and {Oguri}, Masamune and {Shirasaki}, Masato and {Sato}, Masanori},
        title = "{Scatter and bias in weak lensing selected clusters}",
      journal = {\mnras},
         year = 2012,
        month = sep,
       volume = {425},
       number = {3},
        pages = {2287-2298},
          doi = {10.1111/j.1365-2966.2012.21582.x},
archivePrefix = {arXiv},
       eprint = {1204.6117},
 primaryClass = {astro-ph.CO},
       adsurl = {https://ui.adsabs.harvard.edu/abs/2012MNRAS.425.2287H}
}

@ARTICLE{Hamana_2020,
       author = {{Hamana}, Takashi and {Shirasaki}, Masato and {Lin}, Yen-Ting},
        title = "{Weak-lensing clusters from HSC survey first-year data: Mitigating the dilution effect of foreground and cluster-member galaxies}",
      journal = {\pasj},
         year = 2020,
        month = oct,
       volume = {72},
       number = {5},
          eid = {78},
        pages = {78},
          doi = {10.1093/pasj/psaa068},
archivePrefix = {arXiv},
       eprint = {2004.00170},
 primaryClass = {astro-ph.CO},
       adsurl = {https://ui.adsabs.harvard.edu/abs/2020PASJ...72...78H}
}

@ARTICLE{hennawi&spergel,
       author = {{Hennawi}, Joseph F. and {Spergel}, David N.},
        title = "{Shear-selected Cluster Cosmology: Tomography and Optimal Filtering}",
      journal = {\apj},
         year = 2005,
        month = may,
       volume = {624},
       number = {1},
        pages = {59-79},
          doi = {10.1086/428749},
archivePrefix = {arXiv},
       eprint = {astro-ph/0404349},
 primaryClass = {astro-ph},
       adsurl = {https://ui.adsabs.harvard.edu/abs/2005ApJ...624...59H}
}

@ARTICLE{Kaiser_Squires_1993,
       author = {{Kaiser}, Nick and {Squires}, Gordon},
        title = "{Mapping the Dark Matter with Weak Gravitational Lensing}",
      journal = {\apj},
         year = 1993,
        month = feb,
       volume = {404},
        pages = {441},
          doi = {10.1086/172297},
       adsurl = {https://ui.adsabs.harvard.edu/abs/1993ApJ...404..441K}
}

@ARTICLE{Klypin_2016,
       author = {{Klypin}, Anatoly and {Yepes}, Gustavo and {Gottl{\"o}ber}, Stefan and {Prada}, Francisco and {He{\ss}}, Steffen},
        title = "{MultiDark simulations: the story of dark matter halo concentrations and density profiles}",
      journal = {\mnras},
         year = 2016,
        month = apr,
       volume = {457},
       number = {4},
        pages = {4340-4359},
          doi = {10.1093/mnras/stw248},
archivePrefix = {arXiv},
       eprint = {1411.4001},
 primaryClass = {astro-ph.CO},
       adsurl = {https://ui.adsabs.harvard.edu/abs/2016MNRAS.457.4340K}
}

@ARTICLE{Kneib_Natarajan_2011,
       author = {{Kneib}, Jean-Paul and {Natarajan}, Priyamvada},
        title = "{Cluster lenses}",
      journal = {\aapr},
         year = 2011,
        month = nov,
       volume = {19},
          eid = {47},
        pages = {47},
          doi = {10.1007/s00159-011-0047-3},
archivePrefix = {arXiv},
       eprint = {1202.0185},
 primaryClass = {astro-ph.CO},
       adsurl = {https://ui.adsabs.harvard.edu/abs/2011A&ARv..19...47K}
}

@ARTICLE{Kravtsov_Borgani_2012,
       author = {{Kravtsov}, Andrey V. and {Borgani}, Stefano},
        title = "{Formation of Galaxy Clusters}",
      journal = {\araa},
         year = 2012,
        month = sep,
       volume = {50},
        pages = {353-409},
          doi = {10.1146/annurev-astro-081811-125502},
archivePrefix = {arXiv},
       eprint = {1205.5556},
 primaryClass = {astro-ph.CO},
       adsurl = {https://ui.adsabs.harvard.edu/abs/2012ARA&A..50..353K}
}

@ARTICLE{Laigle_2016,
       author = {{Laigle}, C. and {McCracken}, H.~J. and {Ilbert}, O. and {Hsieh}, B.~C. and {Davidzon}, I. and {Capak}, P. and {Hasinger}, G. and {Silverman}, J.~D. and {Pichon}, C. and {Coupon}, J. and {Aussel}, H. and {Le Borgne}, D. and {Caputi}, K. and {Cassata}, P. and {Chang}, Y. -Y. and {Civano}, F. and {Dunlop}, J. and {Fynbo}, J. and {Kartaltepe}, J.~S. and {Koekemoer}, A. and {Le F{\`e}vre}, O. and {Le Floc'h}, E. and {Leauthaud}, A. and {Lilly}, S. and {Lin}, L. and {Marchesi}, S. and {Milvang-Jensen}, B. and {Salvato}, M. and {Sanders}, D.~B. and {Scoville}, N. and {Smolcic}, V. and {Stockmann}, M. and {Taniguchi}, Y. and {Tasca}, L. and {Toft}, S. and {Vaccari}, Mattia and {Zabl}, J.},
        title = "{The COSMOS2015 Catalog: Exploring the 1 < z < 6 Universe with Half a Million Galaxies}",
      journal = {\apjs},
         year = 2016,
        month = jun,
       volume = {224},
       number = {2},
          eid = {24},
        pages = {24},
          doi = {10.3847/0067-0049/224/2/24},
archivePrefix = {arXiv},
       eprint = {1604.02350},
 primaryClass = {astro-ph.GA},
       adsurl = {https://ui.adsabs.harvard.edu/abs/2016ApJS..224...24L}
}

@ARTICLE{Leauthaud_2007,
       author = {{Leauthaud}, Alexie and {Massey}, Richard and {Kneib}, Jean-Paul and {Rhodes}, Jason and {Johnston}, David E. and {Capak}, Peter and {Heymans}, Catherine and {Ellis}, Richard S. and {Koekemoer}, Anton M. and {Le F{\`e}vre}, Oliver and {Mellier}, Yannick and {R{\'e}fr{\'e}gier}, Alexandre and {Robin}, Annie C. and {Scoville}, Nick and {Tasca}, Lidia and {Taylor}, James E. and {Van Waerbeke}, Ludovic},
        title = "{Weak Gravitational Lensing with COSMOS: Galaxy Selection and Shape Measurements}",
      journal = {\apjs},
         year = 2007,
        month = sep,
       volume = {172},
       number = {1},
        pages = {219-238},
          doi = {10.1086/516598},
archivePrefix = {arXiv},
       eprint = {astro-ph/0702359},
 primaryClass = {astro-ph},
       adsurl = {https://ui.adsabs.harvard.edu/abs/2007ApJS..172..219L}
}

@ARTICLE{Leonard_2012,
       author = {{Leonard}, Adrienne and {Pires}, Sandrine and {Starck}, Jean-Luc},
        title = "{Fast calculation of the weak lensing aperture mass statistic}",
      journal = {\mnras},
         year = 2012,
        month = jul,
       volume = {423},
       number = {4},
        pages = {3405-3412},
          doi = {10.1111/j.1365-2966.2012.21133.x10.1002/asna.19141990405},
archivePrefix = {arXiv},
       eprint = {1204.4293},
 primaryClass = {astro-ph.CO},
       adsurl = {https://ui.adsabs.harvard.edu/abs/2012MNRAS.423.3405L}
}

@ARTICLE{Leroy_2023,
       author = {{Leroy}, G. and {Pires}, S. and {Pratt}, G.~W. and {Giocoli}, C.},
        title = "{Fast multi-scale galaxy cluster detection with weak lensing: Towards a mass-selected sample}",
      journal = {\aap},
         year = 2023,
        month = oct,
       volume = {678},
          eid = {A125},
        pages = {A125},
          doi = {10.1051/0004-6361/202346510},
archivePrefix = {arXiv},
       eprint = {2304.01812},
 primaryClass = {astro-ph.CO},
       adsurl = {https://ui.adsabs.harvard.edu/abs/2023A&A...678A.125L}
}

@ARTICLE{Lewis_2000,
       author = {{Lewis}, Antony and {Challinor}, Anthony and {Lasenby}, Anthony},
        title = "{Efficient Computation of Cosmic Microwave Background Anisotropies in Closed Friedmann-Robertson-Walker Models}",
      journal = {\apj},
         year = 2000,
        month = aug,
       volume = {538},
       number = {2},
        pages = {473-476},
          doi = {10.1086/309179},
archivePrefix = {arXiv},
       eprint = {astro-ph/9911177},
 primaryClass = {astro-ph},
       adsurl = {https://ui.adsabs.harvard.edu/abs/2000ApJ...538..473L}
}

@ARTICLE{Maturi_2005,
       author = {{Maturi}, M. and {Meneghetti}, M. and {Bartelmann}, M. and {Dolag}, K. and {Moscardini}, L.},
        title = "{An optimal filter for the detection of galaxy clusters through weak lensing}",
      journal = {\aap},
         year = 2005,
        month = nov,
       volume = {442},
       number = {3},
        pages = {851-860},
          doi = {10.1051/0004-6361:20042600},
archivePrefix = {arXiv},
       eprint = {astro-ph/0412604},
 primaryClass = {astro-ph},
       adsurl = {https://ui.adsabs.harvard.edu/abs/2005A&A...442..851M}
}

@ARTICLE{Miyazaki_2002,
       author = {{Miyazaki}, Satoshi and {Hamana}, Takashi and {Shimasaku}, Kazuhiro and {Furusawa}, Hisanori and {Doi}, Mamoru and {Hamabe}, Masaru and {Imi}, Katsumi and {Kimura}, Masahiko and {Komiyama}, Yutaka and {Nakata}, Fumiaki and {Okada}, Norio and {Okamura}, Sadanori and {Ouchi}, Masami and {Sekiguchi}, Maki and {Yagi}, Masafumi and {Yasuda}, Naoki},
        title = "{Searching for Dark Matter Halos in the Suprime-Cam 2 Square Degree Field}",
      journal = {\apjl},
         year = 2002,
        month = dec,
       volume = {580},
       number = {2},
        pages = {L97-L100},
          doi = {10.1086/345613},
archivePrefix = {arXiv},
       eprint = {astro-ph/0210441},
 primaryClass = {astro-ph},
       adsurl = {https://ui.adsabs.harvard.edu/abs/2002ApJ...580L..97M}
}

@ARTICLE{Miyazaki_2018,
       author = {{Miyazaki}, Satoshi and {Oguri}, Masamune and {Hamana}, Takashi and {Shirasaki}, Masato and {Koike}, Michitaro and {Komiyama}, Yutaka and {Umetsu}, Keiichi and {Utsumi}, Yousuke and {Okabe}, Nobuhiro and {More}, Surhud and {Medezinski}, Elinor and {Lin}, Yen-Ting and {Miyatake}, Hironao and {Murayama}, Hitoshi and {Ota}, Naomi and {Mitsuishi}, Ikuyuki},
        title = "{A large sample of shear-selected clusters from the Hyper Suprime-Cam Subaru Strategic Program S16A Wide field mass maps}",
      journal = {\pasj},
         year = 2018,
        month = jan,
       volume = {70},
          eid = {S27},
        pages = {S27},
          doi = {10.1093/pasj/psx120},
archivePrefix = {arXiv},
       eprint = {1802.10290},
 primaryClass = {astro-ph.CO},
       adsurl = {https://ui.adsabs.harvard.edu/abs/2018PASJ...70S..27M}
}

@ARTICLE{Navarro_1997,
       author = {{Navarro}, Julio F. and {Frenk}, Carlos S. and {White}, Simon D.~M.},
        title = "{A Universal Density Profile from Hierarchical Clustering}",
      journal = {\apj},
         year = 1997,
        month = dec,
       volume = {490},
       number = {2},
        pages = {493-508},
          doi = {10.1086/304888},
archivePrefix = {arXiv},
       eprint = {astro-ph/9611107},
 primaryClass = {astro-ph},
       adsurl = {https://ui.adsabs.harvard.edu/abs/1997ApJ...490..493N}
}

@ARTICLE{Navarro_2004,
       author = {{Navarro}, J.~F. and {Hayashi}, E. and {Power}, C. and {Jenkins}, A.~R. and {Frenk}, C.~S. and {White}, S.~D.~M. and {Springel}, V. and {Stadel}, J. and {Quinn}, T.~R.},
        title = "{The inner structure of {\ensuremath{\Lambda}}CDM haloes - III. Universality and asymptotic slopes}",
      journal = {\mnras},
         year = 2004,
        month = apr,
       volume = {349},
       number = {3},
        pages = {1039-1051},
          doi = {10.1111/j.1365-2966.2004.07586.x},
archivePrefix = {arXiv},
       eprint = {astro-ph/0311231},
 primaryClass = {astro-ph},
       adsurl = {https://ui.adsabs.harvard.edu/abs/2004MNRAS.349.1039N}
}

@ARTICLE{Oguri_2021,
       author = {{Oguri}, Masamune and {Miyazaki}, Satoshi and {Li}, Xiangchong and {Luo}, Wentao and {Mitsuishi}, Ikuyuki and {Miyatake}, Hironao and {More}, Surhud and {Nishizawa}, Atsushi J. and {Okabe}, Nobuhiro and {Ota}, Naomi and {Plazas Malag{\'o}n}, Andr{\'e}s A. and {Utsumi}, Yousuke},
        title = "{Hundreds of weak lensing shear-selected clusters from the Hyper Suprime-Cam Subaru Strategic Program S19A data}",
      journal = {\pasj},
         year = 2021,
        month = aug,
       volume = {73},
       number = {4},
        pages = {817-829},
          doi = {10.1093/pasj/psab047},
archivePrefix = {arXiv},
       eprint = {2103.15016},
 primaryClass = {astro-ph.CO},
       adsurl = {https://ui.adsabs.harvard.edu/abs/2021PASJ...73..817O}
}

@ARTICLE{Parimbelli_2022,
       author = {{Parimbelli}, G. and {Carbone}, C. and {Bel}, J. and {Bose}, B. and {Calabrese}, M. and {Carella}, E. and {Zennaro}, M.},
        title = "{DEMNUni: comparing nonlinear power spectra prescriptions in the presence of massive neutrinos and dynamical dark energy}",
      journal = {\jcap},
         year = 2022,
        month = nov,
       volume = {2022},
       number = {11},
          eid = {041},
        pages = {041},
          doi = {10.1088/1475-7516/2022/11/041},
archivePrefix = {arXiv},
       eprint = {2207.13677},
 primaryClass = {astro-ph.CO},
       adsurl = {https://ui.adsabs.harvard.edu/abs/2022JCAP...11..041P}
}

@ARTICLE{Planck_2016,
       author = {{Planck Collaboration} and {Ade}, P.~A.~R. and {Aghanim}, N. and {Arnaud}, M. and {Ashdown}, M. and {Aumont}, J. and {Baccigalupi}, C. and {Banday}, A.~J. and {Barreiro}, R.~B. and {Barrena}, R. and {Bartlett}, J.~G. and {Bartolo}, N. and {Battaner}, E. and {Battye}, R. and {Benabed}, K. and {Beno{\^\i}t}, A. and {Benoit-L{\'e}vy}, A. and {Bernard}, J. -P. and {Bersanelli}, M. and {Bielewicz}, P. and {Bikmaev}, I. and {B{\"o}hringer}, H. and {Bonaldi}, A. and {Bonavera}, L. and {Bond}, J.~R. and {Borrill}, J. and {Bouchet}, F.~R. and {Bucher}, M. and {Burenin}, R. and {Burigana}, C. and {Butler}, R.~C. and {Calabrese}, E. and {Cardoso}, J. -F. and {Carvalho}, P. and {Catalano}, A. and {Challinor}, A. and {Chamballu}, A. and {Chary}, R. -R. and {Chiang}, H.~C. and {Chon}, G. and {Christensen}, P.~R. and {Clements}, D.~L. and {Colombi}, S. and {Colombo}, L.~P.~L. and {Combet}, C. and {Comis}, B. and {Couchot}, F. and {Coulais}, A. and {Crill}, B.~P. and {Curto}, A. and {Cuttaia}, F. and {Dahle}, H. and {Danese}, L. and {Davies}, R.~D. and {Davis}, R.~J. and {de Bernardis}, P. and {de Rosa}, A. and {de Zotti}, G. and {Delabrouille}, J. and {D{\'e}sert}, F. -X. and {Dickinson}, C. and {Diego}, J.~M. and {Dolag}, K. and {Dole}, H. and {Donzelli}, S. and {Dor{\'e}}, O. and {Douspis}, M. and {Ducout}, A. and {Dupac}, X. and {Efstathiou}, G. and {Eisenhardt}, P.~R.~M. and {Elsner}, F. and {En{\ss}lin}, T.~A. and {Eriksen}, H.~K. and {Falgarone}, E. and {Fergusson}, J. and {Feroz}, F. and {Ferragamo}, A. and {Finelli}, F. and {Forni}, O. and {Frailis}, M. and {Fraisse}, A.~A. and {Franceschi}, E. and {Frejsel}, A. and {Galeotta}, S. and {Galli}, S. and {Ganga}, K. and {G{\'e}nova-Santos}, R.~T. and {Giard}, M. and {Giraud-H{\'e}raud}, Y. and {Gjerl{\o}w}, E. and {Gonz{\'a}lez-Nuevo}, J. and {G{\'o}rski}, K.~M. and {Grainge}, K.~J.~B. and {Gratton}, S. and {Gregorio}, A. and {Gruppuso}, A. and {Gudmundsson}, J.~E. and {Hansen}, F.~K. and {Hanson}, D. and {Harrison}, D.~L. and {Hempel}, A. and {Henrot-Versill{\'e}}, S. and {Hern{\'a}ndez-Monteagudo}, C. and {Herranz}, D. and {Hildebrandt}, S.~R. and {Hivon}, E. and {Hobson}, M. and {Holmes}, W.~A. and {Hornstrup}, A. and {Hovest}, W. and {Huffenberger}, K.~M. and {Hurier}, G. and {Jaffe}, A.~H. and {Jaffe}, T.~R. and {Jin}, T. and {Jones}, W.~C. and {Juvela}, M. and {Keih{\"a}nen}, E. and {Keskitalo}, R. and {Khamitov}, I. and {Kisner}, T.~S. and {Kneissl}, R. and {Knoche}, J. and {Kunz}, M. and {Kurki-Suonio}, H. and {Lagache}, G. and {Lamarre}, J. -M. and {Lasenby}, A. and {Lattanzi}, M. and {Lawrence}, C.~R. and {Leonardi}, R. and {Lesgourgues}, J. and {Levrier}, F. and {Liguori}, M. and {Lilje}, P.~B. and {Linden-V{\o}rnle}, M. and {L{\'o}pez-Caniego}, M. and {Lubin}, P.~M. and {Mac{\'\i}as-P{\'e}rez}, J.~F. and {Maggio}, G. and {Maino}, D. and {Mak}, D.~S.~Y. and {Mandolesi}, N. and {Mangilli}, A. and {Martin}, P.~G. and {Mart{\'\i}nez-Gonz{\'a}lez}, E. and {Masi}, S. and {Matarrese}, S. and {Mazzotta}, P. and {McGehee}, P. and {Mei}, S. and {Melchiorri}, A. and {Melin}, J. -B. and {Mendes}, L. and {Mennella}, A. and {Migliaccio}, M. and {Mitra}, S. and {Miville-Desch{\^e}nes}, M. -A. and {Moneti}, A. and {Montier}, L. and {Morgante}, G. and {Mortlock}, D. and {Moss}, A. and {Munshi}, D. and {Murphy}, J.~A. and {Naselsky}, P. and {Nastasi}, A. and {Nati}, F. and {Natoli}, P. and {Netterfield}, C.~B. and {N{\o}rgaard-Nielsen}, H.~U. and {Noviello}, F. and {Novikov}, D. and {Novikov}, I. and {Olamaie}, M. and {Oxborrow}, C.~A. and {Paci}, F. and {Pagano}, L. and {Pajot}, F. and {Paoletti}, D. and {Pasian}, F. and {Patanchon}, G. and {Pearson}, T.~J. and {Perdereau}, O. and {Perotto}, L. and {Perrott}, Y.~C. and {Perrotta}, F. and {Pettorino}, V. and {Piacentini}, F. and {Piat}, M. and {Pierpaoli}, E. and {Pietrobon}, D. and {Plaszczynski}, S. and {Pointecouteau}, E. and {Polenta}, G. and {Pratt}, G.~W. and {Pr{\'e}zeau}, G. and {Prunet}, S. and {Puget}, J. -L.},
        title = "{Planck 2015 results. XXVII. The second Planck catalogue of Sunyaev-Zeldovich sources}",
      journal = {\aap},
         year = 2016,
        month = sep,
       volume = {594},
          eid = {A27},
        pages = {A27},
          doi = {10.1051/0004-6361/201525823},
archivePrefix = {arXiv},
       eprint = {1502.01598},
 primaryClass = {astro-ph.CO},
       adsurl = {https://ui.adsabs.harvard.edu/abs/2016A&A...594A..27P}
}

@ARTICLE{Pratt_2019,
       author = {{Pratt}, G.~W. and {Arnaud}, M. and {Biviano}, A. and {Eckert}, D. and {Ettori}, S. and {Nagai}, D. and {Okabe}, N. and {Reiprich}, T.~H.},
        title = "{The Galaxy Cluster Mass Scale and Its Impact on Cosmological Constraints from the Cluster Population}",
      journal = {\ssr},
         year = 2019,
        month = feb,
       volume = {215},
       number = {2},
          eid = {25},
        pages = {25},
          doi = {10.1007/s11214-019-0591-0},
archivePrefix = {arXiv},
       eprint = {1902.10837},
 primaryClass = {astro-ph.CO},
       adsurl = {https://ui.adsabs.harvard.edu/abs/2019SSRv..215...25P}
}

@ARTICLE{Schafer_2012,
       author = {{Sch{\"a}fer}, Bj{\"o}rn Malte and {Heisenberg}, Lavinia and {Kalovidouris}, Angelos F. and {Bacon}, David J.},
        title = "{On the validity of the Born approximation for weak cosmic flexions}",
      journal = {\mnras},
         year = 2012,
        month = feb,
       volume = {420},
       number = {1},
        pages = {455-467},
          doi = {10.1111/j.1365-2966.2011.20051.x},
archivePrefix = {arXiv},
       eprint = {1101.4769},
 primaryClass = {astro-ph.CO},
       adsurl = {https://ui.adsabs.harvard.edu/abs/2012MNRAS.420..455S}
}

@ARTICLE{Schirmer_2004,
       author = {{Schirmer}, M. and {Erben}, T. and {Schneider}, P. and {Wolf}, C. and {Meisenheimer}, K.},
        title = "{GaBoDS: The Garching-Bonn Deep Survey. II. Confirmation of EIS cluster candidates by weak gravitational lensing}",
      journal = {\aap},
         year = 2004,
        month = jun,
       volume = {420},
        pages = {75-78},
          doi = {10.1051/0004-6361:20041072},
archivePrefix = {arXiv},
       eprint = {astro-ph/0401203},
 primaryClass = {astro-ph},
       adsurl = {https://ui.adsabs.harvard.edu/abs/2004A&A...420...75S}
}

@ARTICLE{Schneider_1996,
       author = {{Schneider}, Peter},
        title = "{Detection of (dark) matter concentrations via weak gravitational lensing}",
      journal = {\mnras},
         year = 1996,
        month = dec,
       volume = {283},
       number = {3},
        pages = {837-853},
          doi = {10.1093/mnras/283.3.837},
archivePrefix = {arXiv},
       eprint = {astro-ph/9601039},
 primaryClass = {astro-ph},
       adsurl = {https://ui.adsabs.harvard.edu/abs/1996MNRAS.283..837S}
}

@ARTICLE{Schneider_1998,
       author = {{Schneider}, Peter and {van Waerbeke}, Ludovic and {Jain}, Bhuvnesh and {Kruse}, Guido},
        title = "{A new measure for cosmic shear}",
      journal = {\mnras},
         year = 1998,
        month = jun,
       volume = {296},
       number = {4},
        pages = {873-892},
          doi = {10.1046/j.1365-8711.1998.01422.x},
archivePrefix = {arXiv},
       eprint = {astro-ph/9708143},
 primaryClass = {astro-ph},
       adsurl = {https://ui.adsabs.harvard.edu/abs/1998MNRAS.296..873S}
}

@ARTICLE{Schrabback_2018,
       author = {{Schrabback}, T. and {Applegate}, D. and {Dietrich}, J.~P. and {Hoekstra}, H. and {Bocquet}, S. and {Gonzalez}, A.~H. and {von der Linden}, A. and {McDonald}, M. and {Morrison}, C.~B. and {Raihan}, S.~F. and {Allen}, S.~W. and {Bayliss}, M. and {Benson}, B.~A. and {Bleem}, L.~E. and {Chiu}, I. and {Desai}, S. and {Foley}, R.~J. and {de Haan}, T. and {High}, F.~W. and {Hilbert}, S. and {Mantz}, A.~B. and {Massey}, R. and {Mohr}, J. and {Reichardt}, C.~L. and {Saro}, A. and {Simon}, P. and {Stern}, C. and {Stubbs}, C.~W. and {Zenteno}, A.},
        title = "{Cluster mass calibration at high redshift: HST weak lensing analysis of 13 distant galaxy clusters from the South Pole Telescope Sunyaev-Zel'dovich Survey}",
      journal = {\mnras},
         year = 2018,
        month = feb,
       volume = {474},
       number = {2},
        pages = {2635-2678},
          doi = {10.1093/mnras/stx2666},
archivePrefix = {arXiv},
       eprint = {1611.03866},
 primaryClass = {astro-ph.CO},
       adsurl = {https://ui.adsabs.harvard.edu/abs/2018MNRAS.474.2635S}
}

@ARTICLE{Seitz&Schneider_1997,
       author = {{Seitz}, C. and {Schneider}, P.},
        title = "{Steps towards nonlinear cluster inversion through gravitational distortions. III. Including a redshift distribution of the sources.}",
      journal = {\aap},
         year = 1997,
        month = feb,
       volume = {318},
        pages = {687-699},
          doi = {10.48550/arXiv.astro-ph/9601079},
archivePrefix = {arXiv},
       eprint = {astro-ph/9601079},
 primaryClass = {astro-ph},
       adsurl = {https://ui.adsabs.harvard.edu/abs/1997A&A...318..687S}
}

@ARTICLE{Shin_2025,
       author = {{Shin}, T. and {Baxter}, E.~J. and {Lee}, E. and {Battaglia}, N. and {Alarcon}, A. and {Amon}, A. and {Becker}, M. and {Bernstein}, G. and {Bond}, J.~R. and {Campos}, A. and {Chang}, C. and {Chen}, R. and {Choi}, A. and {DeRose}, J. and {Dodelson}, S. and {Doux}, C. and {Dunkley}, J. and {Elvin-Poole}, J. and {Esteves}, J.~H. and {Everett}, S. and {Fert{\'e}}, A. and {Gatti}, M. and {Grandis}, S. and {Gruen}, D. and {Harrison}, I. and {Hill}, J.~C. and {Hilton}, M. and {Jarvis}, M. and {MacCrann}, N. and {McCullough}, J. and {Moodley}, K. and {Mroczkowski}, T. and {Myles}, J. and {Navarro Alsina}, A. and {Nicola}, A. and {Page}, L. and {Pandey}, S. and {Prat}, J. and {Raveri}, M. and {Ried Guachalla}, B. and {Rollins}, R.~P. and {Sanchez}, C. and {Secco}, L.~F. and {Sheldon}, E. and {Sif{\'o}n}, C. and {Troxel}, M. and {Tutusaus}, I. and {von der Linden}, A. and {Wollack}, E. and {Yin}, B. and {Aguena}, M. and {Allam}, S.~S. and {Alves}, O. and {Andrade-Oliveira}, F. and {Bacon}, D. and {Bocquet}, S. and {Brooks}, D. and {Camilleri}, R. and {Carnero Rosell}, A. and {Carretero}, J. and {Castander}, F.~J. and {Costanzi}, M. and {da Costa}, L. and {da Silva Pereira}, M.~E. and {Davis}, T. and {De Vicente}, J. and {Desai}, S. and {Flaugher}, B. and {Frieman}, J. and {Garcia-Bellido}, J. and {Gutierrez}, G. and {Hinton}, S. and {Hollowood}, D.~L. and {Huterer}, D. and {James}, D. and {Lee}, S. and {Marshall}, J. and {Mena-Fern{\'a}ndez}, J. and {Menanteau}, F. and {Miquel}, R. and {Mohr}, J. and {Muir}, J. and {Ogando}, R. and {Plazas Malag{\'o}n}, A. and {Porredon}, A. and {Romer}, K. and {Sanchez}, E. and {Sanchez Cid}, D. and {Sevilla}, I. and {Smith}, M. and {Soares-Santos}, M. and {Suchyta}, E. and {Swanson}, M. and {To}, C. and {Weaverdyck}, N. and {Weller}, J.},
        title = "{Weak Lensing Mass Calibration of the ACT DR5 Galaxy Clusters with the DES Year 3 Weak Lensing Data}",
      journal = {arXiv e-prints},
         year = 2025,
        month = dec,
          eid = {arXiv:2512.18935},
        pages = {arXiv:2512.18935},
          doi = {10.48550/arXiv.2512.18935},
archivePrefix = {arXiv},
       eprint = {2512.18935},
 primaryClass = {astro-ph.CO},
       adsurl = {https://ui.adsabs.harvard.edu/abs/2025arXiv251218935S}
}

@ARTICLE{Springel_2001,
       author = {{Springel}, Volker and {White}, Simon D.~M. and {Tormen}, Giuseppe and {Kauffmann}, Guinevere},
        title = "{Populating a cluster of galaxies - I. Results at z=0}",
      journal = {\mnras},
         year = 2001,
        month = dec,
       volume = {328},
       number = {3},
        pages = {726-750},
          doi = {10.1046/j.1365-8711.2001.04912.x},
archivePrefix = {arXiv},
       eprint = {astro-ph/0012055},
 primaryClass = {astro-ph},
       adsurl = {https://ui.adsabs.harvard.edu/abs/2001MNRAS.328..726S}
}

@book{Starck_1998,
author = {Starck, Jean-Luc and Murtagh, Fionn and Bijaoui, Albert},
year = {1998},
month = {01},
pages = {},
title = {Image Processing and Data Analysis. The Multiscale Approach},
volume = {94},
isbn = {978-0-521-59914-6},
journal = {Journal of the American Statistical Association},
doi = {10.1017/CBO9780511564352}
}

@ARTICLE{Starck_2006,
       author = {{Starck}, J. -L. and {Pires}, S. and {R{\'e}fr{\'e}gier}, A.},
        title = "{Weak lensing mass reconstruction using wavelets}",
      journal = {\aap},
         year = 2006,
        month = jun,
       volume = {451},
       number = {3},
        pages = {1139-1150},
          doi = {10.1051/0004-6361:20052997},
archivePrefix = {arXiv},
       eprint = {astro-ph/0503373},
 primaryClass = {astro-ph},
       adsurl = {https://ui.adsabs.harvard.edu/abs/2006A&A...451.1139S}
}

@ARTICLE{Trobbiani_2025,
       author = {{Trobbiani}, L. and {Maturi}, M. and {Giocoli}, C. and {Moscardini}, L. and {Panebianco}, G.},
        title = "{AMICO-WL: an optimal filtering algorithm for galaxy cluster detections with weak lensing}",
      journal = {\aap},
         year = 2025,
        month = jul,
       volume = {699},
          eid = {A275},
        pages = {A275},
          doi = {10.1051/0004-6361/202553889},
archivePrefix = {arXiv},
       eprint = {2501.16420},
 primaryClass = {astro-ph.CO},
       adsurl = {https://ui.adsabs.harvard.edu/abs/2025A&A...699A.275T}
}

@ARTICLE{Umetsu_review_2020,
       author = {{Umetsu}, Keiichi},
        title = "{Cluster-galaxy weak lensing}",
      journal = {\aapr},
         year = 2020,
        month = dec,
       volume = {28},
       number = {1},
          eid = {7},
        pages = {7},
          doi = {10.1007/s00159-020-00129-w},
archivePrefix = {arXiv},
       eprint = {2007.00506},
 primaryClass = {astro-ph.CO},
       adsurl = {https://ui.adsabs.harvard.edu/abs/2020A&ARv..28....7U}
}

@ARTICLE{Wittman_2001,
       author = {{Wittman}, D. and {Tyson}, J.~A. and {Margoniner}, V.~E. and {Cohen}, J.~G. and {Dell'Antonio}, I.~P.},
        title = "{Discovery of a Galaxy Cluster via Weak Lensing}",
      journal = {\apjl},
         year = 2001,
        month = aug,
       volume = {557},
       number = {2},
        pages = {L89-L92},
          doi = {10.1086/323173},
archivePrefix = {arXiv},
       eprint = {astro-ph/0104094},
 primaryClass = {astro-ph},
       adsurl = {https://ui.adsabs.harvard.edu/abs/2001ApJ...557L..89W}
}

@ARTICLE{Wright&Brainerd_2000,
       author = {{Wright}, Candace Oaxaca and {Brainerd}, Tereasa G.},
        title = "{Gravitational Lensing by NFW Halos}",
      journal = {\apj},
         year = 2000,
        month = may,
       volume = {534},
       number = {1},
        pages = {34-40},
          doi = {10.1086/308744},
       adsurl = {https://ui.adsabs.harvard.edu/abs/2000ApJ...534...34W}
}

@ARTICLE{Wu_2022,
       author = {{Wu}, Hao-Yi and {Costanzi}, Matteo and {To}, Chun-Hao and {Salcedo}, Andr{\'e}s N. and {Weinberg}, David H. and {Annis}, James and {Bocquet}, Sebastian and {da Silva Pereira}, Maria Elidaiana and {DeRose}, Joseph and {Esteves}, Johnny and {Farahi}, Arya and {Grandis}, Sebastian and {Rozo}, Eduardo and {Rykoff}, Eli S. and {Varga}, Tam{\'a}s N. and {Wechsler}, Risa H. and {Zeng}, Chenxiao and {Zhang}, Yuanyuan and {Zhang}, Zhuowen and {DES Collaboration}},
        title = "{Optical selection bias and projection effects in stacked galaxy cluster weak lensing}",
      journal = {\mnras},
         year = 2022,
        month = sep,
       volume = {515},
       number = {3},
        pages = {4471-4486},
          doi = {10.1093/mnras/stac2048},
archivePrefix = {arXiv},
       eprint = {2203.05416},
 primaryClass = {astro-ph.CO},
       adsurl = {https://ui.adsabs.harvard.edu/abs/2022MNRAS.515.4471W}
}

@ARTICLE{Zhou_2024,
       author = {{Zhou}, Conghao and {Wu}, Hao-Yi and {Salcedo}, Andr{\'e}s N. and {Grandis}, Sebastian and {Jeltema}, Tesla and {Leauthaud}, Alexie and {Costanzi}, Matteo and {Sunayama}, Tomomi and {Weinberg}, David H. and {Zhang}, Tianyu and {Rozo}, Eduardo and {To}, Chun-Hao and {Bocquet}, Sebastian and {Varga}, Tamas and {Kwiecien}, Matthew},
        title = "{Forecasting the constraints on optical selection bias and projection effects of galaxy cluster lensing with multiwavelength data}",
      journal = {\prd},
         year = 2024,
        month = nov,
       volume = {110},
       number = {10},
          eid = {103508},
        pages = {103508},
          doi = {10.1103/PhysRevD.110.103508},
archivePrefix = {arXiv},
       eprint = {2312.11789},
 primaryClass = {astro-ph.CO},
       adsurl = {https://ui.adsabs.harvard.edu/abs/2024PhRvD.110j3508Z}
}

@ARTICLE{parimbelli_21,
       author = {{Parimbelli}, G. and {Anselmi}, S. and {Viel}, M. and {Carbone}, C. and {Villaescusa-Navarro}, F. and {Corasaniti}, P.~S. and {Rasera}, Y. and {Sheth}, R. and {Starkman}, G.~D. and {Zehavi}, I.},
        title = "{The effects of massive neutrinos on the linear point of the correlation function}",
      journal = {\jcap},
         year = 2021,
        month = jan,
       volume = {1},
       number = {1},
          eid = {009},
        pages = {009},
          doi = {10.1088/1475-7516/2021/01/009},
archivePrefix = {arXiv},
       eprint = {2007.10345},
 primaryClass = {astro-ph.CO},
       adsurl = {https://ui.adsabs.harvard.edu/abs/2021JCAP...01..009P}
}

@ARTICLE{baratta_22,
       author = {{Baratta}, Philippe and {Bel}, Julien and {Gouyou Beauchamps}, Sylvain and {Carbone}, Carmelita},
        title = "{COVMOS: A new Monte Carlo approach for galaxy clustering analysis}",
      journal = {\aap},
         year = 2023,
        month = may,
       volume = {673},
          eid = {A1},
        pages = {A1},
          doi = {10.1051/0004-6361/202245683},
archivePrefix = {arXiv},
       eprint = {2211.13590},
 primaryClass = {astro-ph.CO},
       adsurl = {https://ui.adsabs.harvard.edu/abs/2023A&A...673A...1B}
}

@ARTICLE{gouyou_25,
       author = {{Gouyou Beauchamps}, S. and {Baratta}, P. and {Escoffier}, S. and {Gillard}, W. and {Bel}, J. and {Bautista}, J. and {Carbone}, C.},
        title = "{Cosmological inference including massive neutrinos from the matter power spectrum: Biases induced by uncertainties in the covariance matrix}",
      journal = {\aap},
         year = 2025,
        month = jan,
       volume = {693},
          eid = {A226},
        pages = {A226},
          doi = {10.1051/0004-6361/202347164},
archivePrefix = {arXiv},
       eprint = {2306.05988},
 primaryClass = {astro-ph.CO},
       adsurl = {https://ui.adsabs.harvard.edu/abs/2025A&A...693A.226G}
}

@ARTICLE{Bel_etal_2025,
       author = {{Euclid Collaboration} and {Bel}, J. and {Gouyou Beauchamps}, S. and {Baratta}, P. and {Blot}, L. and {Carbone}, C. and {Corasaniti}, P.-S. and {Sefusatti}, E. and {Escoffier}, S. and {Gillard}, W. and {Amara}, A. and {Andreon}, S. and {Auricchio}, N. and {Baccigalupi}, C. and {Baldi}, M. and {Bardelli}, S. and {Battaglia}, P. and {Biviano}, A. and {Branchini}, E. and {Brescia}, M. and {Brinchmann}, J. and {Camera}, S. and {Ca{\~n}as-Herrera}, G. and {Capobianco}, V. and {Cardone}, V.~F. and {Carretero}, J. and {Casas}, S. and {Castellano}, M. and {Castignani}, G. and {Cavuoti}, S. and {Chambers}, K.~C. and {Cimatti}, A. and {Colodro-Conde}, C. and {Congedo}, G. and {Conselice}, C.~J. and {Conversi}, L. and {Copin}, Y. and {Costille}, A. and {Courbin}, F. and {Courtois}, H.~M. and {Da Silva}, A. and {Degaudenzi}, H. and {de la Torre}, S. and {De Lucia}, G. and {Dubath}, F. and {Duncan}, C.~A.~J. and {Dupac}, X. and {Farina}, M. and {Farinelli}, R. and {Faustini}, F. and {Ferriol}, S. and {Finelli}, F. and {Fourmanoit}, N. and {Frailis}, M. and {Franceschi}, E. and {Fumana}, M. and {Galeotta}, S. and {George}, K. and {Gillis}, B. and {Giocoli}, C. and {Gracia-Carpio}, J. and {Grazian}, A. and {Grupp}, F. and {Guzzo}, L. and {Haugan}, S.~V.~H. and {Holmes}, W. and {Hormuth}, F. and {Hornstrup}, A. and {Jahnke}, K. and {Jhabvala}, M. and {Joachimi}, B. and {Keih{\"a}nen}, E. and {Kermiche}, S. and {Kubik}, B. and {Kunz}, M. and {Kurki-Suonio}, H. and {Le Brun}, A.~M.~C. and {Ligori}, S. and {Lilje}, P.~B. and {Lindholm}, V. and {Lloro}, I. and {Mainetti}, G. and {Maino}, D. and {Maiorano}, E. and {Mansutti}, O. and {Marggraf}, O. and {Markovic}, K. and {Martinelli}, M. and {Martinet}, N. and {Marulli}, F. and {Massey}, R. and {Medinaceli}, E. and {Mellier}, Y. and {Meneghetti}, M. and {Merlin}, E. and {Meylan}, G. and {Mora}, A. and {Moresco}, M. and {Moscardini}, L. and {Neissner}, C. and {Niemi}, S.-M. and {Padilla}, C. and {Paltani}, S. and {Pasian}, F. and {Pedersen}, K. and {Percival}, W.~J. and {Pettorino}, V. and {Pires}, S. and {Polenta}, G. and {Poncet}, M. and {Popa}, L.~A. and {Raison}, F. and {Renzi}, A. and {Rhodes}, J. and {Riccio}, G. and {Rizzo}, F. and {Romelli}, E. and {Roncarelli}, M. and {Saglia}, R. and {Sakr}, Z. and {S{\'a}nchez}, A.~G. and {Sapone}, D. and {Sartoris}, B. and {Schneider}, P. and {Schrabback}, T. and {Scodeggio}, M. and {Secroun}, A. and {Seidel}, G. and {Seiffert}, M. and {Serrano}, S. and {Simon}, P. and {Sirignano}, C. and {Sirri}, G. and {Stanco}, L. and {Steinwagner}, J. and {Tallada-Cresp{\'\i}}, P. and {Taylor}, A.~N. and {Tereno}, I. and {Tessore}, N. and {Toft}, S. and {Toledo-Moreo}, R. and {Torradeflot}, F. and {Tutusaus}, I. and {Valenziano}, L. and {Valiviita}, J. and {Vassallo}, T. and {Veropalumbo}, A. and {Wang}, Y. and {Weller}, J. and {Zamorani}, G. and {Zucca}, E. and {Ballardini}, M. and {Bozzo}, E. and {Burigana}, C. and {Cabanac}, R. and {Calabrese}, M. and {Di Ferdinando}, D. and {Escartin Vigo}, J.~A. and {Gabarra}, L. and {Mart{\'\i}n-Fleitas}, J. and {Matthew}, S. and {Mauri}, N. and {Metcalf}, R.~B. and {Pezzotta}, A. and {P{\"o}ntinen}, M. and {Porciani}, C. and {Risso}, I. and {Scottez}, V. and {Sereno}, M. and {Tenti}, M. and {Viel}, M. and {Wiesmann}, M. and {Akrami}, Y. and {Alvi}, S. and {Andika}, I.~T. and {Anselmi}, S. and {Archidiacono}, M. and {Atrio-Barandela}, F. and {Bertacca}, D. and {Bethermin}, M. and {Blanchard}, A. and {Borgani}, S. and {Brown}, M.~L. and {Bruton}, S. and {Calabro}, A. and {Camacho Quevedo}, B. and {Caro}, F. and {Carvalho}, C.~S. and {Castro}, T. and {Cogato}, F. and {Conseil}, S. and {Contarini}, S. and {Cooray}, A.~R. and {Davini}, S. and {Desprez}, G. and {D{\'\i}az-S{\'a}nchez}, A. and {Diaz}, J.~J. and {Di Domizio}, S. and {Diego}, J.~M. and {Enia}, A.},
        title = "{Euclid preparation: LXXXVII. Non-Gaussianity of 2-point statistics likelihood: Precise analysis of the matter power spectrum distribution}",
      journal = {arXiv e-prints},
         year = 2025,
        month = nov,
          eid = {arXiv:2511.08266},
        pages = {arXiv:2511.08266},
          doi = {10.48550/arXiv.2511.08266},
archivePrefix = {arXiv},
       eprint = {2511.08266},
 primaryClass = {astro-ph.CO},
       adsurl = {https://ui.adsabs.harvard.edu/abs/2025arXiv251108266E}
}

@ARTICLE{Ingoglia_etal_2025,
       author = {{Euclid Collaboration} and {Ingoglia}, L. and {Sereno}, M. and {Farrens}, S. and {Giocoli}, C. and {Baumont}, L. and {Lesci}, G.~F. and {Moscardini}, L. and {Murray}, C. and {Vannier}, M. and {Biviano}, A. and {Carbone}, C. and {Covone}, G. and {Despali}, G. and {Maturi}, M. and {Maurogordato}, S. and {Meneghetti}, M. and {Radovich}, M. and {Altieri}, B. and {Amara}, A. and {Andreon}, S. and {Auricchio}, N. and {Baccigalupi}, C. and {Baldi}, M. and {Bardelli}, S. and {Bellagamba}, F. and {Bender}, R. and {Bernardeau}, F. and {Bonino}, D. and {Branchini}, E. and {Brescia}, M. and {Brinchmann}, J. and {Camera}, S. and {Capobianco}, V. and {Carretero}, J. and {Casas}, S. and {Castellano}, M. and {Castignani}, G. and {Cavuoti}, S. and {Cimatti}, A. and {Colodro-Conde}, C. and {Congedo}, G. and {Conselice}, C.~J. and {Conversi}, L. and {Copin}, Y. and {Courbin}, F. and {Courtois}, H.~M. and {Cropper}, M. and {da Silva}, A. and {Degaudenzi}, H. and {de}, Lucia G. and {Dinis}, J. and {Dubath}, F. and {Duncan}, C.~A.~J. and {Dupac}, X. and {Dusini}, S. and {Ealet}, A. and {Farina}, M. and {Faustini}, F. and {Ferriol}, S. and {Fosalba}, P. and {Frailis}, M. and {Franceschi}, E. and {Fumana}, M. and {Galeotta}, S. and {Gillard}, W. and {Gillis}, B. and {Gomez-Alvarez}, P. and {Grazian}, A. and {Grupp}, F. and {Guzzo}, L. and {Haugan}, S.~V.~H. and {Holmes}, W. and {Hormuth}, F. and {Hornstrup}, A. and {Hudelot}, P. and {Ilic}, S. and {Jahnke}, K. and {Jhabvala}, M. and {Joachimi}, B. and {Keihanen}, E. and {Kermiche}, S. and {Kiessling}, A. and {Kilbinger}, M. and {Kubik}, B. and {Kummel}, M. and {Kunz}, M. and {Kurki-Suonio}, H. and {Ligori}, S. and {Lilje}, P.~B. and {Lindholm}, V. and {Lloro}, I. and {Mainetti}, G. and {Maiorano}, E. and {Mansutti}, O. and {Marcin}, S. and {Marggraf}, O. and {Markovic}, K. and {Martinelli}, M. and {Martinet}, N. and {Marulli}, F. and {Massey}, R. and {Medinaceli}, E. and {Mei}, S. and {Melchior}, M. and {Mellier}, Y. and {Merlin}, E. and {Meylan}, G. and {Moresco}, M. and {Munari}, E. and {Niemi}, S.-M. and {Padilla}, C. and {Paech}, K. and {Paltani}, S. and {Pasian}, F. and {Pedersen}, K. and {Percival}, W.~J. and {Pettorino}, V. and {Pires}, S. and {Polenta}, G. and {Poncet}, M. and {Popa}, L.~A. and {Pozzetti}, L. and {Raison}, F. and {Renzi}, A. and {Rhodes}, J. and {Riccio}, G. and {Romelli}, E. and {Roncarelli}, M. and {Rossetti}, E. and {Saglia}, R. and {Sakr}, Z. and {Salvignol}, J.-C. and {Sanchez}, A.~G. and {Sapone}, D. and {Sartoris}, B. and {Schirmer}, M. and {Schneider}, P. and {Secroun}, A. and {Seidel}, G. and {Serrano}, S. and {Sirignano}, C. and {Sirri}, G. and {Stanco}, L. and {Steinwagner}, J. and {Tallada-Crespi}, P. and {Tavagnacco}, D. and {Taylor}, A.~N. and {Tereno}, I. and {Toledo-Moreo}, R. and {Torradeflot}, F. and {Tutusaus}, I. and {Valenziano}, L. and {Vassallo}, T. and {Verdoes Kleijn}, G. and {Veropalumbo}, A. and {Wang}, Y. and {Weller}, J. and {Zamorani}, G. and {Zucca}, E. and {Bolzonella}, M. and {Bozzo}, E. and {Burigana}, C. and {Calabrese}, M. and {di Ferdinando}, D. and {Escartin Vigo}, J.~A. and {Farinelli}, R. and {Finelli}, F. and {Gracia-Carpio}, J. and {Matthew}, S. and {Pezzotta}, A. and {Pontinen}, M. and {Scottez}, V. and {Tenti}, M. and {Viel}, M. and {Wiesmann}, M. and {Akrami}, Y. and {Allevato}, V. and {Anselmi}, S. and {Archidiacono}, M. and {Atrio-Barandela}, F. and {Ballardini}, M. and {Bertacca}, D. and {Bethermin}, M. and {Blanchard}, A. and {Blot}, L. and {Bohringer}, H. and {Borgani}, S. and {Bruton}, S. and {Cabanac}, R. and {Calabro}, A. and {Canas-Herrera}, G. and {Cappi}, A. and {Caro}, F. and {Carvalho}, C.~S. and {Castro}, T. and {Chambers}, K.~C. and {Contarini}, S. and {Cooray}, A.~R. and {Costanzi}, M.},
        title = "{Euclid preparation: LXV. Determining the weak lensing mass accuracy and precision for galaxy clusters}",
      journal = {\aap},
         year = 2025,
        month = apr,
       volume = {695},
          eid = {A280},
        pages = {A280},
          doi = {10.1051/0004-6361/202452122},
archivePrefix = {arXiv},
       eprint = {2409.02783},
 primaryClass = {astro-ph.CO},
       adsurl = {https://ui.adsabs.harvard.edu/abs/2025A&A...695A.280E}
}

\appendix

\section{Modelling NFW haloes in a convergence field accounting for a source redshift distribution}
\label{sec:appendix:nfw_mocks}
The NFW \citep{Navarro_1997} parameterization was shown to represent DM density profiles well in the Cold Dark Matter (CDM) paradigm across a wide range of halo masses \citep{Navarro_2004, Klypin_2016}. For a given overdensity $\Delta$, the NFW profile reads

\begin{equation}
    \rho_{\rm NFW}(\theta) = \frac{\rho_{s}}{\frac{c_{\Delta,c} \theta}{R_{\Delta,c}} \left(1+ \frac{c_{\Delta,c} \theta}{R_{\Delta,c}}\right)^2}.
\label{eq:nfw}
\end{equation}

\noindent Here $R_{\Delta}$ is the radius $\theta$ at which $\bar\rho(\theta) = \Delta \rho_{\rm crit}$ and $c_\Delta$ is the concentration parameter, $c_\Delta=R_{\Delta} / r_s$ with $r_s$ the scale radius (defined as the radius at which the NFW slope is equal to $-2$). The scaled density $\rho_{s}$ can be reformulated as a function of the critical density of the Universe $\rho_c(z)$ as

\begin{equation}
    \rho_{s} = \frac{\Delta}{3} \frac{c_\Delta^3}{\ln(1+c_\Delta)-\frac{c_\Delta}{1+c_\Delta}}\rho_c(z_l).
\label{eq:rhoc}
\end{equation}

Computing the projected surface mass density $\Sigma(\theta)$ from a given density profile $\rho(\theta)$ (Eq. \ref{eq:sigma}) holds an analytical solution in the NFW case (in a more general case, it can still be obtained through numerical integration). We used the Python package \begin{tt}Colossus\end{tt}\footnote{\url{https://bdiemer.bitbucket.io/colossus}} \citep{Diemer_2018} to model $\Sigma(\theta)$ by taking as inputs the halo masses from the N-body halo catalogue and a fixed concentration of 4. Finally, by using the corresponding cluster redshifts $z_l$, we accounted for the redshift distribution of the sources $p(z_s)$ by computing the expected $\kappa(\theta)$ profile as an average over $p(z_s)$. Specifically, following \citet{Seitz&Schneider_1997, Bartelmann&Schneider_2001, Berge_2010}, we evaluate:

\begin{equation}
    \langle\kappa(\theta)\rangle  = \int_{z_l}^{\infty} \kappa (\theta, z_s)p(z_s)dz_s = \Sigma(\theta) \int_{z_l}^{\infty} \frac{p(z_s)}{\Sigma_{\mathrm{crit}}(z_l, z_s)}dz_s,
    \label{eq:kappa_zdist}
\end{equation}
ensuring the expected convergence is sensitive to the real distribution of source galaxies behind each lens redshift. 
Note that in our work, the convergence maps will generally contain foreground sources ($z_s < z_l$), this dilution effect is accounted for by setting the lensing efficiency $\Sigma_{\mathrm{crit}}^{-1}(z_s)$ to 0 when $z_s<z_l$ \citep{giocoli17}.

\section{Theoretical lensing S/N}
\label{sec:appendix:S/N}
The expected S/N for any overdensity in a circular aperture in the convergence $\kappa_E$ is given by \citep{Berge_2010}:

\begin{equation}
    \nu = \frac{\sqrt{n_g}}{\sigma_{e}}\sqrt{\int d^2 \theta \kappa(\theta)^2},
    \label{eq:kappa_S/N}
\end{equation}
where $n_g$ is the galaxy density, $\sigma_\epsilon$ the shape noise, and $\theta$ is the radial distance to the centre.

In the case where the convergence field is convolved with any given filter $U$, we can generalize equation \ref{eq:kappa_S/N} to the following:

\begin{equation}
    \nu_U = \frac{\sqrt{n_g}}{\sigma_{e}}\frac{\int d^2 \theta \kappa(\theta)U( \theta )}{\sqrt{\int d^2 \theta U^2( \theta )}}.
    \label{eq:filter_S/N}
\end{equation}

We can replace $U$ by the expression of our wavelet filters $U_{W_i}$ (Eq. \ref{eq:wvlt_filters}) to compute the theoretical wavelet dependent S/N:
\begin{equation}
    \nu_i = \frac{\sqrt{n_g}}{\sigma_{e}}\frac{\int d^2 \theta \kappa(\theta)U_{W_i}( \theta )}{\sqrt{\int d^2 \theta U^2_{W_i}( \theta )}}.
    \label{eq:wvlt_S/N}
\end{equation}

In the NFW framework, we can model the expected convergence signal using Eq. \ref{eq:kappa_zdist} and the procedure described in Appendix \ref{sec:appendix:nfw_mocks}. Finally, we compute the expected S/N in each scale by injecting the modelled convergence into Eq. \ref{eq:wvlt_S/N}.

\onecolumn

\section{Complementary figures}
\begin{figure*}[htbp]
    \centering
    \includegraphics[width=\linewidth, trim={0cm 0cm 0cm 0cm},clip]{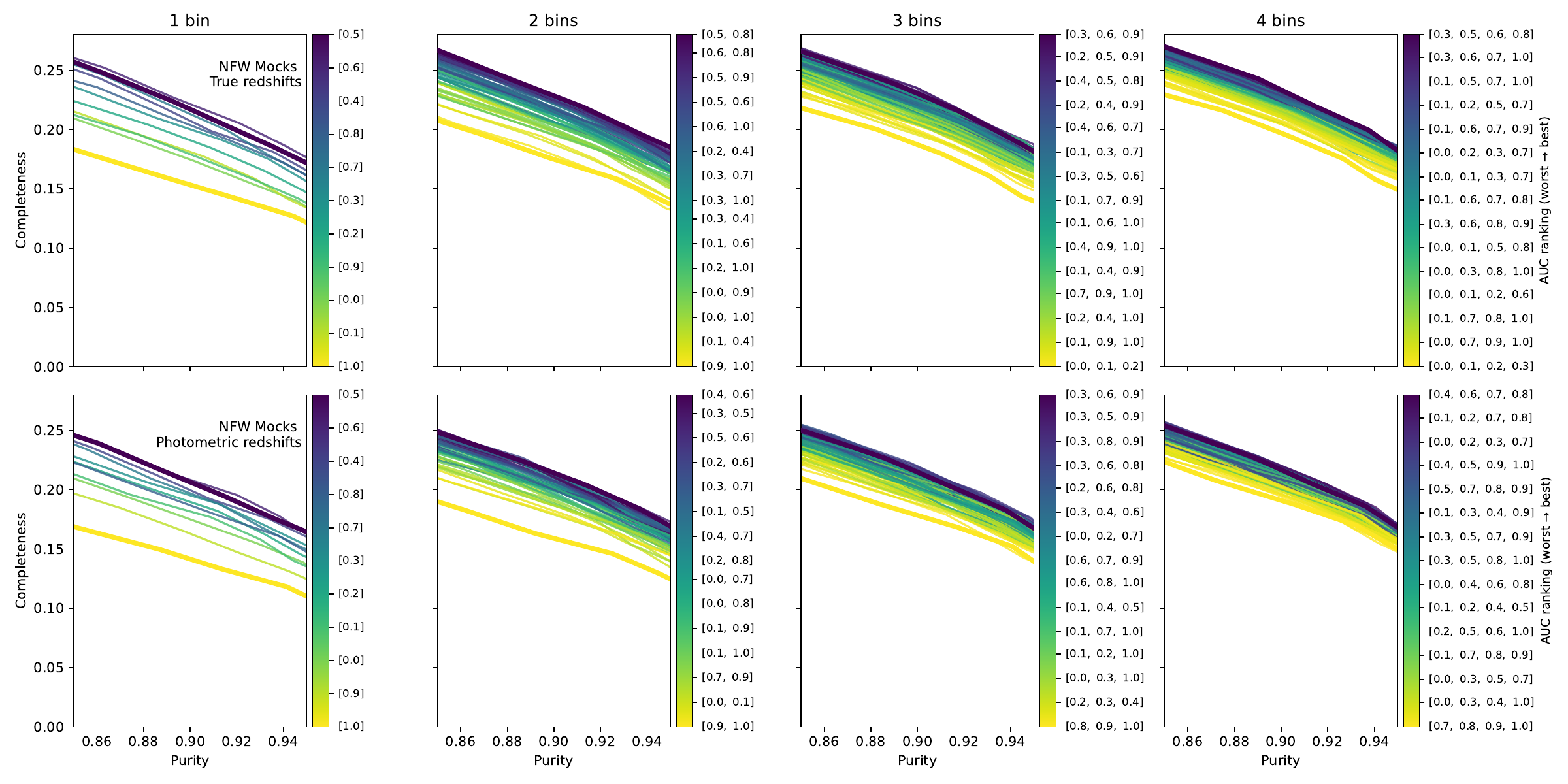}
    \caption{Purity-completeness plots for LSS-free NFW mocks. All source redshift bins combinations are represented in the 1-, 2-, 3-, and 4-bins scenarios (from left to right). Curves are ranked by AUC (colour bar). In multi-bin cases, not all combinations figure on the colour bars.\\Top row: using true source redshift distribution. \\Bottom row: using source redshift with $Euclid$-like photometric errors.}
    \label{fig:CvsP_nfw_mocks_all}
\end{figure*}

\begin{figure*}[htbp]
    \centering
    \includegraphics[width=\linewidth, trim={0cm 0cm 0cm 0cm},clip]{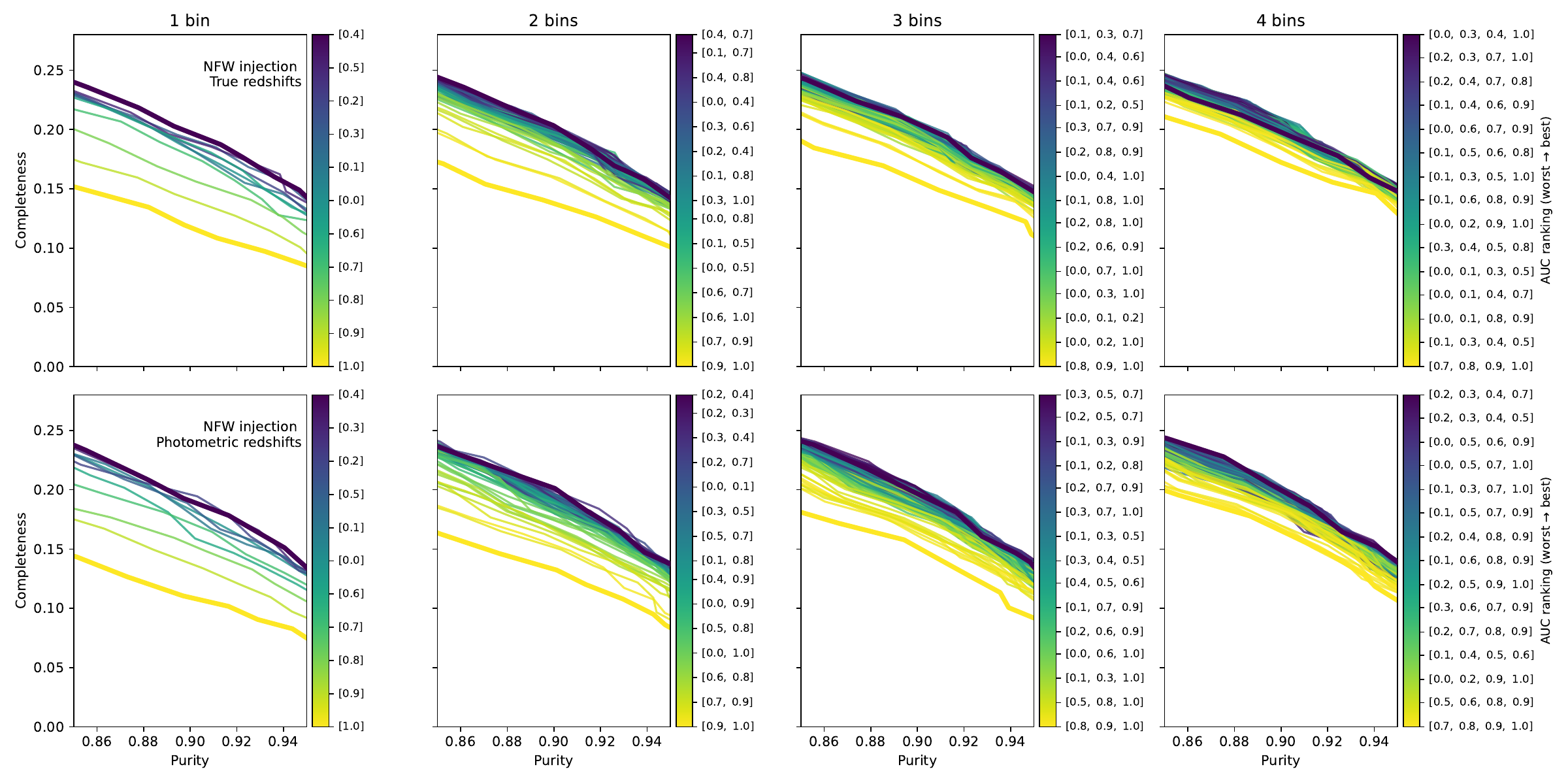}
    \caption{Purity-completeness plots for injected NFW haloes in N-body fields. All source redshift bins combinations are represented in the 1-, 2-, 3-, and 4-bins scenarios (from left to right). Curves are ranked by AUC (colour bar). In multi-bin cases, not all combinations figure on the colour bars.\\Top row: using true source redshift distribution. \\Bottom row: using source redshift with $Euclid$-like photometric errors.}
    \label{fig:CvsP_nfw_inj_all}
\end{figure*}

\begin{figure*}[htbp]
    \centering
    \includegraphics[width=\linewidth, trim={0cm 0cm 0cm 0cm},clip]{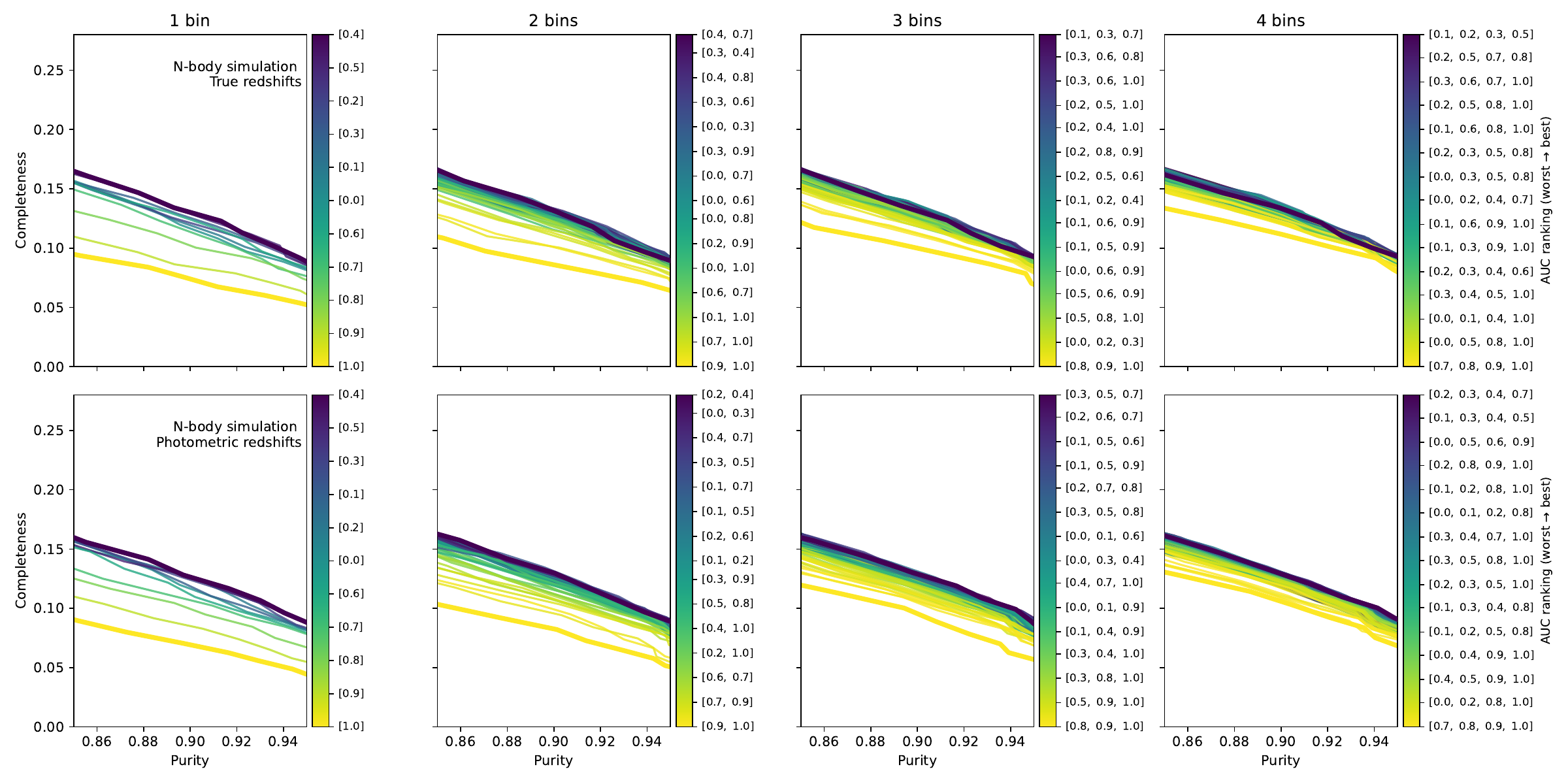}
    \caption{Purity-completeness plots on the N-body fields. All source redshift bins combinations are represented in the 1-, 2-, 3-, and 4-bins scenarios (from left to right). Curves are ranked by AUC (colour bar). In multi-bin cases, not all combinations figure on the colour bars.\\Top row: using true source redshift distribution. \\Bottom row: using source redshift with $Euclid$-like photometric errors.}
    \label{fig:CvsP_challenge_all}
\end{figure*}

\begin{figure*}[htbp]
    \centering
    \includegraphics[width=\linewidth, trim={0cm 0cm 0cm 0cm},clip]{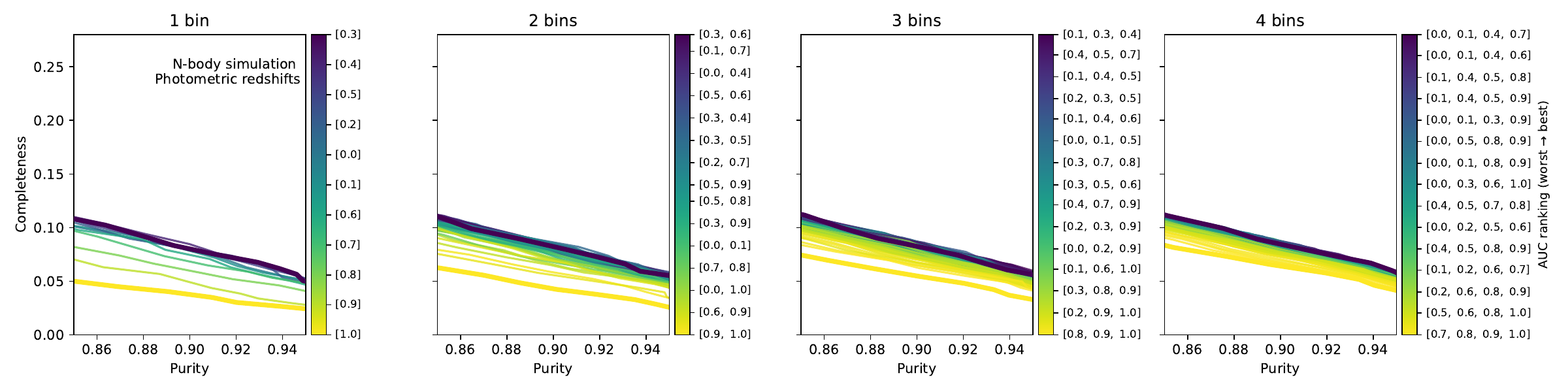}
    \caption{Purity-completeness plots on the N-body fields in the single scale analysis (W3) using source redshift with $Euclid$-like photometric errors. All source redshift bins combinations are represented in the 1-, 2-, 3-, and 4-bins scenarios (from left to right). Curves are ranked by AUC (colour bar). In multi-bin cases, not all combinations figure on the colour bars. }
    \label{fig:CvsP_challenge_all_w3}
\end{figure*}

\end{document}